\documentstyle[12pt,epsf,axodraw]{article}
\newcommand{\Tint}[1]{{\hbox{$\sum$}\!\!\!\!\!\!\int}_{\!\!\!\!#1}}
\newcommand{\la}[1]{\label{#1}}
 
\newcommand{\be}{\begin{equation}}
\newcommand{\ee}{\end{equation}}
\newcommand{\ba}{\begin{eqnarray}}
\newcommand{\ea}{\end{eqnarray}}
\newcommand{\bi}{\begin{itemize}}
\newcommand{\ei}{\end{itemize}}
\newcommand{\rmi}[1]{{\mbox{\scriptsize #1}}}
\newcommand{\nr}[1]{(\ref{#1})}
\newcommand{\tr}{{\rm Tr\,}}
\newcommand{\re}{{\rm Re\,}}
\newcommand{\Hc}{{\rm H.c.\ }}
\newcommand{\im}{\mathop{\rm Im}}
\newcommand{\nn}{\nonumber \\}
\newcommand{\fr}[2]{{\frac{#1}{#2}}}
\newcommand{\msbar}{\overline{\mbox{\rm MS}}}

\newcommand{\muT}{\overline\mu_T}
\newcommand{\<}{\langle} 
\renewcommand{\>}{\rangle}  

\newcommand{\tb}{\tan\!\beta\,}
\renewcommand{\sb}{\sin\!\beta\,}
\newcommand{\cb}{\cos\!\beta\,}
\newcommand{\ctb}{\cot\!\beta\,}
\newcommand{\sbb}{\sin\!2\beta\,}
\newcommand{\cbb}{\cos\!2\beta\,}
\newcommand{\ssb}{\sin^2\!\beta\,}
\newcommand{\ccb}{\cos^2\!\beta\,}
\newcommand{\pf}{\frac{1}{16\pi^2}}
\newcommand{\htt}{\frac{h_t^2}{16\pi^2}}
\newcommand{\gpf}{\frac{g^2}{16\pi^2}}
\newcommand{\htttt}{\frac{h_t^4}{16\pi^2}}
\newcommand{\gghtt}{\frac{g^2h_t^2}{16\pi^2}}
\newcommand{\gggg}{\frac{g^4}{16\pi^2}}

\newcommand{\bmu}{\bar{\mu}}
\newcommand{\Lf}{\ln\frac{(4\bmu)^2}{\muT^2}}
\newcommand{\Lb}{\ln\frac{\bmu^2}{\muT^2}}
\newcommand{\lnmQ}{\ln\frac{\bmu^2}{m_Q^2}}
\newcommand{\lnmQQ}{\biggl(\ln\frac{\bmu^2}{m_Q^2}+1\biggr)}
\newcommand{\lnmQQQ}{\biggl(\ln\frac{\bmu^2}{m_Q^2}+2\biggr)}
\newcommand{\Att}{|\hat A_t|^2}
\newcommand{\muu}{|\hat\mu|^2}
\newcommand{\pint}{\int\! dp}
\newcommand{\pslash}{\slash\!\!\! p}
\newcommand{\lnQT}{\ln\frac{m_Q}{\muT}}
\newcommand{\lnfQT}{\ln\frac{4m_Q}{\muT}}
\newcommand{\lnstQT}{\ln\frac{16m_Q}{\muT}}
\newcommand{\lnQeT}{\ln\frac{m_Q}{e\muT}}
\newcommand{\lnQheT}{\ln\frac{m_Q}{e^{1/2}\muT}}

\newcommand{\eq}{Eq.~}
\newcommand{\eqs}{Eqs.~}

\newcommand{\se}{Sec.~}

\def\lsi{\raise0.3ex\hbox{$<$\kern-0.75em\raise-1.1ex\hbox{$\sim$}}}
\def\gsi{\raise0.3ex\hbox{$>$\kern-0.75em\raise-1.1ex\hbox{$\sim$}}}
\newcommand{\lsim}{\mathop{\lsi}}
\newcommand{\gsim}{\mathop{\gsi}}
\makeatletter \@addtoreset{equation}{section} \makeatother
\renewcommand{\theequation}{\arabic{section}.\arabic{equation}}
 
\begin{document}
 
\begin{titlepage}
\begin{flushright}
CERN-TH/98-364\\
NORDITA-98/69HE\\
hep-ph/9811369\\
\end{flushright}
\begin{centering}
\vfill
 
{\bf HIGGS SECTOR CP-VIOLATION AT THE \\ 
     ELECTROWEAK PHASE TRANSITION}
\vspace{0.8cm}

M. Laine$^{\rm a,b}$\footnote{mikko.laine@cern.ch} and
K. Rummukainen$^{\rm c}$\footnote{kari@nordita.dk}  \\

\vspace{0.3cm}
{\em $^{\rm a}$Theory Division, CERN, CH-1211 Geneva 23,
Switzerland\\}
\vspace{0.3cm}
{\em $^{\rm b}$Department of Physics,
P.O.Box 9, 00014 University of Helsinki, Finland\\}
\vspace{0.3cm}
{\em $^{\rm c}$NORDITA, Blegdamsvej 17,
DK-2100 Copenhagen \O, Denmark}

\vspace{0.7cm}
{\bf Abstract}
 
\end{centering}
 
\vspace{0.3cm}\noindent
We consider explicit and spontaneous CP-violation related to the
profile of the two Higgs doublets at the MSSM electroweak phase
transition.  We find, in accordance with previous results, that in
principle spontaneous CP-violation could exist in the MSSM at finite
temperatures, but when the constraints from experiment and the
strength of the transition are taken into account, the relevant region
of the parameter space is rather restricted. Nevertheless, we show that in
this small region, perturbative estimates need not be reliable, and
non-perturbatively the region might be slightly larger (or smaller). 
To allow for more precise perturbative studies and for a lattice 
study of the non-perturbative corrections, we construct an effective 
3d theory for the light stops and the two Higgses in the regime of 
large $m_Q$. The 3d theory involves CP-violating parameters and allows 
to determine CP-odd observables.
\vfill
\noindent

 
\vspace*{1cm}
 
\noindent
CERN-TH/98-364\\
NORDITA-98/69HE\\
November 1998
 
\vfill

\end{titlepage}
 
\section{Introduction}

The physics problem motivating the studies of the cosmological
electroweak phase transition is that it could 
contribute to the matter-antimatter
(baryon) asymmetry of the Universe~\cite{krs}.  
Several different ingredients are required for 
baryon asymmetry generation:
one needs anomalous baryon number violating processes
in the symmetric high temperature phase, microscopic
C- and CP-violation, and thermal non-equilibrium (for 
a review, see~\cite{rs}). 

Recently, much progress has
been made in understanding the order and strength of the
electroweak phase transition (which determine 
whether there is non-equilibrium and what the rate of 
baryon number violation is in the broken phase), 
and also in understanding the rate of 
baryon number violation in the symmetric phase.
It has been found that
in the Standard Model, there is no electroweak phase
transition at all for the experimentally allowed
Higgs masses (for a review, see~\cite{own}).
In the MSSM, in contrast, 
there can be a first order phase transition, the strength 
of which has been studied perturbatively up to 
2-loop level~\cite{e}--\cite{cm} and, because of the 
potentially bad convergence of the perturbative series, 
non-perturbatively employing dimensional reduction and 
three-dimensional (3d) lattice simulations~\cite{ml}--\cite{mssmsim}. 
The conclusion of these studies was that, from 
the point of view of the strength of the transition,  
Higgs masses up to 105...110 GeV are allowed for
electroweak baryogenesis, provided that the right-handed 
stop is light, $m_{\tilde t_R}<m_\rmi{top}$.

As to the baryon number violation rate, 
its parametric form has been determined analytically~\cite{asy,db}, 
and numerical estimates exist both in the symmetric phase~\cite{moore_sy}
and in the broken phase~\cite{moore_br} of the theory. The latter case 
serves to determine the magnitude of higher order corrections
to the perturbative 1-loop saddle point computation; the corrections
were found to be as small as expected.

However, it seems that the third requirement for
baryon number generation, namely CP-violation, remains less 
well understood. If there were no CP-violation, one would produce
the same amounts of baryons and anti-baryons, and no net
asymmetry would arise. In the Standard Model, there 
is CP-violation as is experimentally observed in the $K^0$-system,
and this phenomenon is explained by the single complex parameter
in the Kobayashi-Maskawa matrix. The Standard Model
CP-violation is probably not large enough for 
the purpose of baryon number generation, though 
(for a review, see~\cite{rs}). 
Hence one would need new sources of CP-violation which,
at the same time,  
do not produce any observable effects violating 
the existing experimental constraints. 

An interesting prospect for such a mechanism
is provided by the MSSM. Indeed, there are new sources
of CP-violation there. Fixing some regularization scale,
one can redefine the fields so that the new CP-violating 
parameters appear in the trilinear couplings of the
two Higgs doublets $H_1, H_2$ and the squarks:
\be
{\cal L}_\rmi{CP}=h_t(A_t^*\tilde H_2-\mu H_1)^\dagger Q_\alpha U_\alpha
+ \Hc, \la{CPparams}
\ee
where $\tilde H_2 = i\sigma_2 H_2^*$, 
$h_t$ is the top Yukawa coupling, $A_t,\mu$ are complex
parameters with the dimension of mass ($A_t$ is induced
by soft supersymmetry breaking while $\mu$ is 
assumed to be the supersymmetric
mass parameter appearing in the Higgs sector), $Q_\alpha$
is the left-handed third generation squark, and $U_\alpha$
is the right-handed third generation squark. However, if the 
CP-violating effects are proportional to the complex
phases of $A_t,\mu$, one is in the regime of explicit CP-violation, 
and then one again has to be careful in order not to violate the 
experimental constraints (for a review, see~\cite{wb})\footnote{The
experimental constraints might be weaker 
than often assumed, though \cite{bgk}.}.

There is also another possibility for CP-violation 
in the MSSM. Indeed, since one has two Higgs doublets in the theory,
there is the possibility of spontaneous CP-violation~\cite{lee}.
This means that even though all the parameters were real, there
could be a dynamically generated phase angle between $H_1,H_2$
(of course, without any explicit CP-violation, an angle and 
minus the angle are equally likely). The possibility 
of spontaneous CP-violation is in principle there 
already at $T=0$, but in practice the parameter 
values needed (in particular, a small CP-odd Higgs mass $m_A$)
are experimentally excluded~\cite{mp}. On the other
hand, it may be that spontaneous CP-violation is more
easily realized at finite temperatures~\cite{emq}, 
or even that it takes place only in 
the phase boundary between the symmetric and broken
phases~\cite{cpr,fkot}. A non-trivial CP-violating
profile in the phase boundary could conceivably be quite useful 
for electroweak baryogenesis~\cite{mstv}--\cite{rio}, 
and at the same time, one would not need to worry about the constraints 
that have to be satisfied in the broken phase at $T=0$. 

The parameter region found in~\cite{emq,cpr,fkot} 
for finite $T$ spontaneous CP-violation consists
of relatively small values of $m_A$ and 
relatively large values of $\tb$. Such values are
not particularly ideal from the point of view of the
strength of the transition, but could still be allowed
in the small stop regime~\cite{cm}.

In this paper, we analyze the general prospects for
finite $T$ spontaneous
CP-violation in some detail. The emphasis is on understanding 
what kind of perturbative and non-perturbative effects there are, 
and whether the non-perturbative effects could be studied with lattice
simulations. As a tool for these discussions, we construct an
effective 3d theory which involves the CP-violating dynamics. 
Finally, we explain how this theory can be used for more precise
perturbative and non-perturbative numerical studies than
carried out here. 

It should be noted that all these finite $T$ studies 
concern the thermodynamical equilibrium situation at $T=T_c$. 
During the actual history
of the electroweak phase transition, the static equilibrium 
situation with $T=T_c$ is only reached if reheating to 
$T_c$ takes place after the transition, which is not 
generically true~\cite{kl,cm}. However, understanding the 
equilibrium situation is clearly a good starting point. 

The plan of the paper is the following. In \se\ref{sec:theory}
we briefly describe the functional form of the relevant 
dimensionally reduced effective 3d theory, and the parametric
magnitude of its couplings. In \se\ref{sec:param}
we study, parametrically, what kinds of ``instabilities''
could in principle occur. Such possible instabilities are 
the deviation of an effective $\tb$-parameter from a constant
value within the phase boundary, 
spontaneous CP-violation, and the breaking of the 
charge neutrality of the vacuum. 
In \se\ref{sec:profile} we discuss 
the profile of $\tb$ in some more detail, in particular with
respect to whether it could be studied with
lattice simulation. In \se\ref{sec:cp} we do 
the same for spontaneous CP-violation. We conclude
in \se\ref{sec:concl}. The details related to the 
derivation of the effective 3d theory described in 
\se\ref{sec:theory} are presented in the Appendix.

\section{The effective 3d theory for $H_1,H_2,U$}
\la{sec:theory}

To discuss the physics of finite temperature
CP-violation, we shall first 
construct the corresponding effective 3d theory, using the 
method of dimensional reduction
applied to systems with phase transitions~\cite{generic}. 
The motivation for
this approach is that it implements automatically 
the resummations needed at finite temperatures and is
thus the simplest way of seeing which infrared (IR) 
problems can be cured and which remain non-perturbative. 
The 3d theory may also allow for lattice simulations, 
as discussed below\footnote{It is interesting to note that
the very good accuracy of dimensional reduction
(in the Standard Model) has recently
been confirmed also fully non-perturbatively with 4d finite 
temperature lattice simulations~\cite{cfh}.}.
 
We work in the light stop regime, so that the 
dynamical degrees of freedom appearing in the 
effective theory are the SU(2) and SU(3) gauge 
fields, the two Higgs doublets $H_1,H_2$, 
and the right-handed stop $U$ (in principle, 
the U(1) gauge field should also be kept there, but
it is expected to induce only small corrections~\cite{su2u1}). 
Moreover, we
assume that the left-handed stop mass parameter
is relatively large, $m_Q\gsim 0.6$ TeV,
whereas the mixing parameters and the other mass 
parameters are not excessively so:  
$|A_t| \lsim m_Q$, and $|\mu|$ and the other mass 
parameters are small compared with $\pi T$.
The details of the derivation of the effective
theory are in the Appendix.
We need here the main features only.

The effective 3d Lagrangian is the most
general gauge-invariant Lagrangian involving the given
degrees of freedom:
\ba
{\cal L}_\rmi{3d} & = & 
\fr14 F^a_{ij}F^a_{ij}+\fr14 G^A_{ij}G^A_{ij} \nn
& + & (D_i^w H_1)^\dagger(D_i^w H_1) + 
(D_i^w H_2)^\dagger(D_i^w H_2) + (D_i^s U)^\dagger (D_i^s U) \nn
& + &  m_1^2(T) H_1^\dagger H_1 +  m_2^2(T) H_2^\dagger H_2 
+ \Bigl[ m_{12}^2(T) H_1^\dagger \tilde H_2 +\Hc  \Bigr] 
+ m_U^2(T) U^\dagger U \nn
& + &  \gamma_1 U^\dagger U H_1^\dagger H_1 + 
 \gamma_2 U^\dagger U H_2^\dagger H_2
 + \Bigl[\gamma_{12} U^\dagger U H_1^\dagger \tilde H_2+\Hc\Bigr] 
 + \lambda_U (U^\dagger U )^2 \nn
& + &  \lambda_1 (H_1^\dagger H_1)^2 + 
 \lambda_2 (H_2^\dagger H_2)^2 + 
 \lambda_3 H_1^\dagger H_1 H_2^\dagger H_2 + 
 \lambda_4 H_1^\dagger \tilde H_2 \tilde H_2^\dagger H_1 \nn
& + & \Bigl[ \lambda_5 (H_1^\dagger \tilde H_2)^2 + 
 \lambda_6 H_1^\dagger H_1 H_1^\dagger \tilde H_2 +
 \lambda_7 H_2^\dagger H_2 H_1^\dagger \tilde H_2 + \Hc\Bigr],
\la{H1H2action} 
\ea
where $D_i^w=\partial_i-i g t^a A_i^a$, 
$D_i^s=\partial_i-i g_{S}T^A C_i^A$ are the 
SU(2) and SU(3) covariant derivatives,
$t^a=\sigma^a/2$ where $\sigma^a$ are the Pauli matrices,
and $U$ denotes the complex conjugate of the original
right-handed stop field appearing in \eqs\nr{CPparams}, \nr{A0act}.
Of the parameters in \eq\nr{H1H2action},  
$m_{12}^2, \gamma_{12}, \lambda_{5}, \lambda_{6}, \lambda_{7}$ 
can in principle be complex. The complex phases are due to
the couplings $A_t,\mu$ in \eq\nr{CPparams}. 
On the other hand, in the case of spontaneous 
CP-violation, all the parameters can be real.

It should be noted that after dimensional reduction, 
the effective theory also contains the zero components of
the gauge fields $A_0^a,C_0^A$, with mass terms 
$(1/2) m_{A_0}^2 A_0^aA_0^a$, $(1/2) m_{C_0}^2 C_0^AC_0^A$, 
and, in principle, all the possible gauge-invariant 
quartic interactions involving $A_0^a,C_0^A$ and the
degrees of freedom in \eq\nr{H1H2action}
(most of these can be constructed out of the
gauge-invariant bilinears $A_0^aA_0^a$, $C_0^AC_0^A$, $H_1^\dagger H_1$, 
$H_2^\dagger H_2$, $H_1^\dagger \tilde H_2$, $U^\dagger U$).
However, as argued in Appendix~\ref{gauge}, these terms
are not essential for the present discussion, and thus
$A_0^a,C_0^A$ can be integrated out.

It is also important to note that all other 
CP-violating operators, such as those related
to $F\tilde F$ (see, e.g., \cite{mstv}), 
result in higher-order operators
whose contributions to Higgs sector CP-violation
are suppressed relative to the effects arising
within the theory in \eq\nr{H1H2action}~\cite{pviol}.

The expressions for the couplings in \eq\nr{H1H2action}
arising from dimensional reduction are summarized
in Appendix~\ref{summary} (see also~\cite{ml,ck,lo,emq,cpr,fkot}). 
Here it is sufficient to know
the leading parametric behaviour of the couplings:
omitting multiplicative numerical factors and 
replacing logarithms by terms of order unity
in the 1-loop terms proportional to $1/(16\pi^2)$, we get
\ba
& & 
m_1^2(T) \sim m_{1}^2 +
\biggl(\fr38 g^2 -\fr14 h_t^2 \muu  \biggr) T^2, \la{aeq1st} \\
& & 
m_2^2(T) \sim m_{2}^2 +
\biggl(\fr38 g^2+\fr12 h_t^2 -\fr14 h_t^2 \Att  \biggr) T^2, \\
& & 
m_{12}^2(T) \sim m_{12}^2  + 
\fr14 h_t^2 \hat A_t^*\hat\mu^* T^2 -
\gpf M_2\mu^* + \htt A_t^*\mu^*, \la{m12} \\
& & 
m_U^2(T) \sim m_{U}^2 + 
\biggl( \fr23 g_S^2 + \fr13 h_t^2 -\fr16 h_t^2 (\Att+\muu)
\biggr) T^2, \\
& & 
\lambda_U \sim \fr16 g_S^2, \quad 
\gamma_1 \sim  -h_t^2 \muu, \quad \la{aeqcoup} 
\gamma_2 \sim 
h_t^2 (1-\Att), \\
& & 
\gamma_{12} \sim 
h_t^2 \hat A_t^* \hat \mu^* - 
\gghtt \biggl[
\frac{m_{12}^2}{(2\pi T)^2}  + \frac{M_2 \mu^*}{(\pi T)^2}\biggr], \\
& & 
\lambda_1 \sim 
\fr18 (g^2+g'^2), \quad 
\lambda_2 \sim 
\fr18 (g^2+g'^2) + \htttt, \\
& &
\lambda_3 \sim 
\fr14 (g^2-g'^2), \quad 
\lambda_4 \sim 
-\fr12 g^2, \\
& & 
\lambda_5 \sim 
\pf \biggl[
h_t^4 (\hat A_t^*\hat \mu^*)^2 - 
g^4 \frac{m_{12}^4}{(2\pi T)^4} + 
g^4 \frac{(M_2\mu^*)^2}{(\pi T)^4}
\biggr], \la{la5} \\
& & 
\lambda_6 \sim 
\pf \biggl[
-g^2 h_t^2 \hat A_t^* \hat\mu^* -
g^4 \frac{m_{12}^2}{(2\pi T)^2} -
g^4 \frac{M_2 \mu^*}{(\pi T)^2} \biggr], \la{la6} \\
& & 
\lambda_7 \sim 
\pf \biggl[
- h_t^4 \hat A_t^* \hat\mu^* 
- g^4 \frac{m_{12}^2}{(2\pi T)^2} -
g^4 \frac{M_2 \mu^*}{(\pi T)^2}
\biggr], \la{la7} \la{aeqlast}
\ea
where $m_1^2,m_2^2,m_{12}^2,m_U^2$ denote the zero-temperature
mass parameters, $M_2$ is the SU(2) gaugino mass parameter,
and
\be
\hat A_t \equiv \frac{A_t}{m_Q},\quad
\hat \mu \equiv \frac{\mu}{m_Q}. \la{Atmudef}
\ee
The signs (which do have 
some significance) have been obtained by assuming that 
$m_Q\sim 1$ TeV, $T\sim 100$ GeV, $|A_t|\sim|\mu|\lsim 100$ GeV;
for the general case, see \eqs\nr{seq1st}--\nr{seqlast}.
It should be noted that loop effects within the
effective theory may generate contributions which are 
larger than those in \eqs\nr{la5}--\nr{aeqlast}; 
see \se\ref{sec:cp}.

We have written in \eqs\nr{aeq1st}--\nr{aeqlast} the couplings
in 4d units, so that there is an overall factor $\int_0^\beta d\tau=1/T$ 
in the action. This factor can, as usual, be defined away by 
a scaling of the fields, and the result is that the quartic
couplings get simply multiplied by $T$.

Note that the form of \eq\nr{H1H2action}
imposes several constraints on the parameters. There are
various kinds of constraints (see, e.g., \cite{clm}). 
First, the quartic part of
the potential should grow (or remain the same) 
in every direction of the field space, for the field
values that the effective theory is applicable.
Otherwise the potential is ``unbounded from below''.
Second, the standard electroweak minimum should be
at least a metastable one at low temperatures. 

Denoting by $v_1,v_2$ the expectation values of the neutral
components of $H_1,H_2$, $\tb=v_2/v_1$, and
\be
\gamma_\rmi{eff} = 
h_t^2 \Bigl[ -\muu \ccb + (1-\Att) \ssb + 
(\hat A_t^* \hat\mu^* + \hat A_t \hat \mu)
\cos\!\beta\sin\!\beta \Bigr],
\ee 
the first constraint says that for {\em all values of} 
$\beta$ that lead to $\gamma_\rmi{eff}<0$, one must require
\be
\gamma_\rmi{eff}^2 < \frac{1}{12} g_S^2 \tilde g^2 \cos^2 2\beta,
\la{ufb}
\ee
where $\tilde g^2 = g^2 + g'^2$.
For instance, taking $\tb=0$, this means that there is 
an instability in the plane $v_2=0$ unless
\be
h_t^4 |\hat\mu|^4 < \frac{\tilde g^2 g_S^2}{12}. 
\ee
Thus, we have to require at least that $|\hat\mu| \ll 0.4$.
There is a similar but weaker constraint for $\hat A_t$, 
$\hat A_t\lsim 1$. 

The second constraint applies in the broken electroweak
phase, for the physical value of $\tb$. Then the simplest 
requirement is just that the right-handed stop mass squared
be positive, which means that for the physical value of $\tb$, 
$\gamma_\rmi{eff}$ must be positive
(for the values of $m_U^2$ we are interested in),
and moreover, that
\be
1-| \hat A_t - \hat \mu^*\cot\!\beta|^2 > -\frac{m_U^2}{m_\rmi{top}^2}.
\la{meta}
\ee
This constraint is typically weaker than the one in \eq\nr{ufb}. 
Another version of \eq\nr{meta} applies at finite temperatures
when one requires that the history of the Early Universe leads
us to the standard electroweak minimum~\cite{bjls,cqw2,cm}. 

In order
to satisfy these constraints, we assume that $\Att,\muu\ll 1$.

\section{Parametric estimates for instabilities,\\
and when is further reduction possible?}
\la{sec:param}

Let us now inspect
the Higgs doublet part of the action in \eq\nr{H1H2action}.
Taking into account the tree-level expressions for the
parameters $m_{1}^2,m_{2}^2$ (see \eqs\nr{seq1st}, \nr{seq2nd}), 
we observe that the sum of the eigenvalues of the Higgs 
mass matrix is 
\be
m_1^2(T) + m_2^2(T) = m_A^2 +
\biggl(\fr34 g^2+\fr12 h_t^2 -\fr14 h_t^2 (\Att+\muu) 
+ {\cal O}(g^3) \biggr) T^2
\sim (gT)^2. \la{summ}
\ee
Here $m_A$ is the CP-odd Higgs mass, 
we use the parametric convention $g\sim h_t\sim g_S$, 
and we assume that $\Att+\muu\ll 1$, see \se\ref{sec:theory}.
On the other hand, at the phase transition point, 
there is a direction along which one of the Higgs doublets 
is light, $m_\rmi{light}^2 \sim (g^2 T)^2$
(this is because the phase transition takes at the tree-level 
place when the second derivative of the effective potential
vanishes in some direction). 
Then the other Higgs
doublet must be heavy, $m_\rmi{heavy}^2 \sim (gT)^2$, 
since the sum of the eigenvalues is constrained by \eq\nr{summ}.
Thus the question arises, how could there be instabilities
such as spontaneous
CP-violation, since one of the Higgs doublets is heavy
and could possibly be integrated out, as done in~\cite{ml,ck,lo}?
If there is just one dynamical Higgs doublet, there cannot
be any spontaneous CP-violation.

To establish rules for power-counting, recall that 
the first order electroweak phase transition is 
generated typically by gauge field (or stop) 1-loop terms 
of the form $-T m_W^3$ (or $-T m_{\tilde t_R}^3$). 
Moreover, let us assume that 
the quartic scalar self-coupling~$\lambda$ is of the parametric
order $g^2$. Then, according to the standard argument, 
the phase transition takes place
when the tree-level quadratic and quartic terms are
of the same order of magnitude as the 1-loop contribution:
\be
g^4 T^2 v^2 \sim g^3 T v^3 \sim g^2 v^4, \la{hmagn}
\ee
where $v$ denotes the light Higgs field expectation value. 
It follows that $v\sim gT$. Note that 
numerically, one would like to have 
in the broken phase that $v\gsim T$ due to the sphaleron 
erasure bound \cite{krs,rs}, but since the bosonic temperature 
scale means really $\sim 2\pi T$, even these values 
can parametrically be thought of as $\sim gT$. The actual 
convergence of the perturbative expansion of course gets better
with an increasing numerical value of $v$.

Thus, we discuss values of order 
$v\sim gT$ below. Let us stress that
these estimates are assumed to hold also within the phase 
boundary, even though strictly speaking a fixed value of $v$
is only obtained in one of the homogeneous (meta)stable phases.

\subsection{Estimates in the diagonalized theory}
\la{sec:diag}

Let us now diagonalize the two Higgs doublet model mass matrix
(consisting of $m_1^2(T)$, $m_2^2(T)$, $m_{12}^2(T)$ 
in \eqs\nr{aeq1st}--\nr{m12}) at the phase transition point.
We denote $m^2 = m_\rmi{light}^2$, $M^2 = m_\rmi{heavy}^2$, 
and the corresponding fields by $h,H$, with vacuum expectation
values $v_h,v_H$ (note that $h,H$ are linear combinations
of $H_1,\tilde H_2$). Then, the diagonalized Higgs potential 
is of the form
\ba
V(h,H) & = & m^2 h^\dagger h + M^2 H^\dagger H +
\lambda_1 (h^\dagger h)^2 + \lambda_2 (H^\dagger H)^2 + 
\lambda_3 h^\dagger h H^\dagger H + 
\lambda_4 h^\dagger H H^\dagger h \nn
& + & \Bigl[
\lambda_5 (h^\dagger H)^2 + 
\lambda_6 h^\dagger h h^\dagger H + 
\lambda_7 H^\dagger H h^\dagger H 
 + \Hc\Bigr]. \la{vorig}
\ea
The couplings $\lambda_1,...,\lambda_7$ are some linear
combinations of those in \eq\nr{H1H2action}
(see, e.g., \eq(6.21) in~\cite{ml}), but for
simplicity we do not introduce a new notation.

Since one would expect that the dynamics is 
dominated by the light field $h$, 
consider the set of Green's functions with only 
$h$ in the outer legs. Conceivably one can, 
order by order in perturbation theory, organize the
contributions of the heavy fields $H$ so that they 
correspond to contributions to the
parameters of an effective theory involving only $h$, 
\be
{\cal L}_\rmi{eff} = 
(D_i^w h)^\dagger (D_i^w h)+
m_\rmi{eff}^2 h^\dagger h + 
\lambda_\rmi{eff} (h^\dagger h)^2 + \ldots, \la{veff}
\ee
where the terms not displayed are higher order operators. 
The expressions for the effective parameters $m_\rmi{eff}^2$, 
$\lambda_\rmi{eff}$ have been discussed 
in~\cite{ml}: e.g., 
\be
\lambda_\rmi{eff} = 
\lambda_1-\frac{T}{8\pi M}
\Bigl(\lambda_3^2+\lambda_3\lambda_4+\fr12 \lambda_4^2+
2|\lambda_5|^2+12|\lambda_6|^2-6\lambda_6^*\lambda_7
-6 \lambda_6\lambda_7^*\Bigr). \la{leff}
\ee
The complication in the derivation
of these parameters is that due
to the interactions proportional to $\lambda_6,\lambda_7$,
which break the $h\to -h,H\to -H$ symmetries, there 
are reducible diagrams contributing to the
effective parameters, related to the mixing between
$h,H$ generated at 1-loop level (see~\cite{ml}).

However, clearly the construction of the theory in \eq\nr{veff}
cannot be valid for arbitrary quartic couplings:
one must require that $\lambda_i T/M\ll 1$
for the loop expansion parameters related to 
corrections such as those in \eq\nr{leff} to be small.
Parametrically, this means that
$\lambda_i\lsim {\cal O}(g^2)$. Now, it is couplings
at most of this order that arise from dimensional reduction, 
see \eqs\nr{aeqcoup}--\nr{aeqlast}.
Let us thus inspect whether this regime is consistent 
with instabilities such as spontaneous CP-violation.

To analyse the kind of instabilities that can appear
in the theory in \eq\nr{H1H2action}, it is illuminating
to write $V(h,H)$ in terms of other sets of variables. Note first
that $V(h,H)$ can be written as a 2nd order
polynomial in the explicitly real and gauge invariant operators
\be
O_1 = h^\dagger h, \quad
O_2 = H^\dagger H, \quad
O_3 = \re h^\dagger H, \quad
O_4 = \im h^\dagger H. \la{ops}
\ee
These operators are constrained by the inequality
$O_3^2 + O_4^2 \le O_1 O_2$, but the existence
of four operators nevertheless shows
that there are, in general, four independent physical
directions in the field space. However, the representation 
of the potential
in terms of the $O_i$'s is not exceedingly useful for 
studying these different
directions, due to the constraint.

For perturbative studies, a more useful representation is 
obtained by using global SU(2) and U(1) transformations to rotate
$h$ to unitary gauge and
to remove one phase angle from $H$. Moreover, 
let us introduce an angular variable $\beta$ by defining
$\tb=v_H/v_h$, $v^2 = v_h^2 + v_H^2$ (note that this $\tb$ 
is not the same as the usual zero-temperature
parameter $\tb=v_2/v_1$ of the MSSM, since
we have already made a rotation to a basis where $m_{12}^2(T)=0$,
to arrive at \eq\nr{vorig}). 
Then, the two doublets can be written as
\be
h = \frac{v}{\sqrt{2}}
\left( 
\begin{array}{l}
\cos\beta \\
0
\end{array}
\right), \quad
H = \frac{v}{\sqrt{2}}
\left( 
\begin{array}{l}
\sin\beta\cos\theta e^{i\phi} \\
\sin\beta\sin\theta
\end{array}
\right). \la{prms}
\ee
In terms of these variables, the 
tree-level potential in \eq\nr{vorig} becomes
\ba
V(h,H) & = & \fr12 v^2 \Bigl(m^2 \ccb + M^2 \ssb \Bigr) \nn
& + & \fr14 v^4 \Bigl\{\lambda_1 \cos^4\!\beta + 
\lambda_2\sin^4\!\beta + \fr14\sin^2\!2\beta
\Bigl[\lambda_3 +\cos^2\theta (\lambda_4 + 
2 \lambda_5 \cos\!2\phi )\Bigr] \nn
& + & \sbb\cos\!\theta \cos\!\phi (\lambda_6 \ccb+\lambda_7\ssb)
\Bigr\}. \la{VhH}
\ea

There are now three possible
types of deviations from the naive case that the 
heavy field $H$ is zero and only the light field $h$ 
is dynamical (i.e., $\beta=0$).
First, $H$ can get a non-vanishing expectation value, 
corresponding to $\beta \neq 0$. 
As we will see, this deviation
generically takes place and does not introduce any qualitative
changes in the thermodynamical 
properties of the system. It turns out that $\beta$ 
typically depends on the value of $v$.  

The other two changes are more drastic in the sense that 
they correspond to genuinely new ``phases''. Both of these
deviations require that $\beta\neq 0$, see \eq\nr{prms}. 
The first possibility is 
that $\theta$ can be  non-zero. This means that the 
doublet $H$ in \eq\nr{prms} has both upper and lower components 
non-vanishing. When gauge fields are taken into account, this 
would mean that the photon becomes massive, and acts thus as a 
(non-local) order parameter for this phase (another way to say it is 
that the vacuum is not neutral). Clearly, the parameters 
are expected to be such that this possibility is not realized.

The second possibility is that $\phi\neq 0$.
This case corresponds to spontaneous CP-violation, if
all the parameters appearing in \eq\nr{VhH} are real. 
Indeed, for the scalar fields, we can define the 
C-transformation as complex conjugation, 
while the P-trans\-for\-mation reverses the signs of the 
spatial coordinates but leaves the scalar fields intact
(for a more general analysis, see~\cite{bgg}). Then 
\be
h \stackrel{\rmi{CP}}{\to} h^*, \quad
H \stackrel{\rmi{CP}}{\to} H^*.
\ee
The action 
corresponding to \eq\nr{H1H2action} is invariant under
this transformation, for real parameters. 
However, the angle $\phi$ and the 
gauge-invariant operator
$O_4$ in \eq\nr{ops} are not invariant.
Let us denote $O_\rmi{CP} = O_4$:
$O_\rmi{CP}$ acts as an order parameter for 
this phase.

\subsubsection*{The implications of $\beta\neq 0$}

Due to the fact that $M^2\sim (gT)^2$ is large, we can assume
all the three kinds of deviations to be small. For 
$\beta$ this will be justified presently; for $\theta$, $\phi$
this assumption is no restriction, since, without a loss
of generality, we are choosing to inspect 
when the origin $\theta=0,\phi=0$ becomes 
unstable (one may need to invert the signs of $\lambda_6,\lambda_7$
for this). 
Expanding to second order in the angles
in \eq\nr{VhH} and noting that 
$\lambda_i v^4\sim g^2 (gT)^4 \ll M^2v^2 \sim (gT)^4$,  
it is easy to see that 
\be
\beta \approx \sin\!\beta \approx -\fr12 \frac{\lambda_6 v^2}{M^2}
\sim {\cal O}(g^2). \la{beold}
\ee
Here we assumed that $\lambda_6\sim {\cal O}(g^2)$, 
$v\sim M\sim gT$.
Thus, indeed, $\beta$ is in general non-vanishing but small, 
and depends on $v$.

Let us now inspect what a value $\beta\neq 0$ means for 
the thermodynamics of the system and, in particular, for
the construction of the effective theory in \eq\nr{veff}, 
which is supposed to account for the thermodynamics.
As seen from \eq\nr{VhH}, 
a non-zero value of $\beta$ causes a shift in the free energy
of the order 
\be
\fr12 M^2 v^2 \beta^2 + \fr12 \lambda_6 v^4 \beta \approx
-\fr18 \frac{\lambda_6^2 v^6}{M^2} \sim 
{\cal O}(g^8T^4). \la{xcontr}
\ee
Is this contribution contained in 
the theory in \eq\nr{veff}? The answer is, no, 
since the term in \eq\nr{xcontr} corresponds to 
a higher-order operator $\sim (h^\dagger h)^3$. 
However, this contribution is suppressed
by the relative amount ${\cal O}(g^2)$
with respect to contributions within the effective theory 
in \eq\nr{veff}, since they are of the order
\be
m_\rmi{eff}^2 v^2 \sim \lambda_\rmi{eff}v^4 \sim {\cal O}(g^6T^4).
\ee 
(The same estimate arises by comparing \eq\nr{xcontr}
directly with the 
tree-level 6-point function generated within the 
effective theory, $\sim (\lambda_\rmi{eff}^2/m_\rmi{eff}^2)v^6$.)
Due to the relative suppression ${\cal O}(g^2)$, 
the negative sign of \eq\nr{xcontr}
is also not a problem in the range of $v$ considered. 
Thus, the effective theory is still useful and accurate
in the phase transition region (for weak coupling), 
and $\beta\neq 0$ does not change
the properties of the system qualitatively.

As a side remark, let us note 
that, for $\lambda_6\sim {\cal O}(g^2)$, 
the accuracy estimate ${\cal O}(g^2)$ above 
differs from the one, ${\cal O}(g^3)$,
in \cite{generic,pviol}. The reason is that with the 
present type of interactions, there can be reducible (left) 
diagrams with a heavy internal line, 
in additional to irreducible ones (right):  
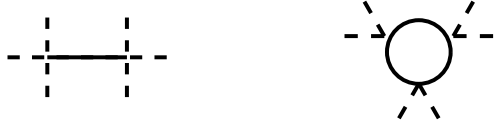
\begin{center}
\begin{picture}(270,60)(0,0)

\SetWidth{1.5}
\SetScale{1.0}
\DashLine(20,30)(80,30){5}
\DashLine(35,15)(35,45){5}
\Line(35,30)(65,30)
\DashLine(65,15)(65,45){5}

\CArc(175,32)(12,0,360)
\DashLine(175,20)(182.5,7){5}
\DashLine(175,20)(167.5,7){5}
\DashLine(188,38)(203,38){5}
\DashLine(188,38)(195.5,51){5}
\DashLine(162,38)(147,38){5}
\DashLine(162,38)(154.5,51){5}

\end{picture}
\end{center}
In the Standard Model case
considered in \cite{generic,pviol}, in contrast, 
there are only irreducible diagrams, whose contribution
is suppressed by one power of $\lambda_3 T/M\sim {\cal O}(g)$
with respect to the reducible ones. In the MSSM,  
$\lambda_6$ is in fact typically parametrically
smaller than ${\cal O}(g^2)$, 
so that even there one may get errors  
smaller than ${\cal O}(g^2)$. 

\subsubsection*{The implications of $\phi,\theta\neq 0$}

Let us then look at the directions $\phi,\theta$. 
To quadratic order, the 
part of the effective potential depending on them is
\be
\frac{1}{4}v^4 \Bigl\{
\beta^2 \Bigl[
\lambda_4 (1-\theta^2) + 2\lambda_5 (1-\theta^2-2\phi^2)
\Bigr] + 2\beta\lambda_6
\Bigl[1-\fr12(\theta^2+\phi^2) \Bigr] \Bigr\}.
\ee
Thus, $\phi$ can be non-zero if
\be
\beta^2 (4\lambda_5) + \beta\lambda_6 > 0, \la{phieq}
\ee
while $\theta$ can be non-zero if
\be
\beta^2 (\lambda_4+ 2\lambda_5)+ \beta\lambda_6 > 0.  \la{thetaeq}
\ee
Suppose that all the couplings $\lambda_i$ are of
the same order of magnitude ($\sim g^2$). Then, the $\beta^2$-
terms in \eqs\nr{phieq}, \nr{thetaeq}
are of order ${\cal O}(g^6)$ since $\beta\sim {\cal O}(g^2)$ 
and can be neglected, and the L.H.S.\ of \eq\nr{phieq} is,
according to \eq\nr{beold}, 
\be
\beta\lambda_6 \approx -\fr12 \frac{\lambda_6^2v^2}{M^2} < 0.
\ee
In other words, for couplings of this 
order of magnitude, the inequality in \eq\nr{phieq}
is never satisfied, $\phi$ is zero, 
and CP is not spontaneously violated!

When can $\phi$ then be non-zero? It is seen 
from \eq\nr{phieq} that one must have $\lambda_5>0$ and  
\be
\lambda_5 > -\frac{\lambda_6}{4\beta} 
\approx \fr12 \frac{M^2}{v^2} \sim {\cal O}(1).
\ee
(Recall that we assume $v\sim M\sim gT$.)
Hence, CP can be spontaneously violated 
only if $\lambda_5$ is of the parametric order ${\cal O}(1)$!
Clearly, in that case the construction of the effective 
theory in \eq\nr{veff} is not valid.

Another possibility is that $\lambda_7$, which did not play 
any role in the previous estimates, is large. 
But to be effective, it has to be at least of the order 
\be
\lambda_7 \sim \frac{\lambda_6}{\beta^2}\sim 
\lambda_6 {\cal O}(g^{-4}).
\ee
If $\lambda_6\sim {\cal O}(g^2)$, $\lambda_7$ needs to be even 
larger than $\lambda_5$, and the reduction in \eq\nr{leff}
again does not work.

It is seen from \eq\nr{thetaeq} that the conditions 
for the breaking of $\theta$ are similar as those for
the breaking of $\phi$, except that 
$\lambda_4 < 0$ can have a further stabilizing effect.

In conclusion, 
CP could in principle be spontaneously violated, 
and the photon could even become massive, if the quartic 
couplings of the two Higgs doublet model are suitable. 
In such a case, further reduction into a single
Higgs doublet model does not work, because some 
of the expansion parameters ${\cal O}(\lambda_i T/M)$ 
would be large. However, this case is not 
a natural consequence of dimensional reduction in the MSSM, 
since the couplings produced this way are of too small a
parametric magnitude, see \eqs\nr{aeqcoup}--\nr{aeqlast}
(this statement is true even after radiative effects
within the effective theory are taken into account, 
see \se\ref{sec:cp}). 
To get spontaneous CP-violation in practice, one
thus needs somewhat large numerical factors and 
suitable parameter values, which overcome the 
parametric suppression. We will study these 
issues in \se\ref{sec:cp}.

\subsection{Estimates in the original theory}
\la{eiot}

In the estimates so far we used for simplicity 
the diagonalized effective theory in \eq\nr{vorig}. 
For completeness, 
it is perhaps good to see how the same estimates arise
in terms of the original variables, before a redefinition 
which allowed us to get rid of $m_{12}^2$. 

Let us start by reviewing the general tree-level condition
for spontaneous CP-viola\-tion within the theory in \eq\nr{H1H2action}
(this discussion corresponds to those in~\cite{emq,cpr}).
Putting now $\theta\to 0$, we choose,
corresponding to \eq\nr{prms},
\be
H_1 = 
\frac{1}{\sqrt{2}}
\left(
\begin{array}{l}
v_1 \\
0
\end{array}
\right), \quad
H_2 = 
\frac{1}{\sqrt{2}}
\left(
\begin{array}{l}
0 \\
v_2 e^{-i\phi}
\end{array}
\right), \quad
H_1^\dagger \tilde H_2 = \fr12 v_1 v_2 e^{i\phi}.
\ee
Then, the part of the potential depending on $\phi$ is 
\be
V(\cos\phi) = 
m_{12}^2(T) v_1v_2 \cos\phi + 
\fr12 \Bigl[\lambda_5 v_1^2v_2^2 ( 2\cos^2 \phi-1 )  + 
\lambda_6 v_1^3 v_2 \cos\phi + 
\lambda_7 v_2^3 v_1 \cos\phi\Bigr].
\ee
Denoting again $\tb=v_2/v_1$,
we note that the second order polynomial 
$V(\cos\phi)$ has a non-trivial minimum, if $\lambda_5 > 0$ and
\be
|f(\beta,v)| = \left|\frac{m_{12}^2(T) 
+ (1/2) (\lambda_6 \ccb+\lambda_7\ssb)v^2}
{2\lambda_5\sb\cb v^2}\right| < 1, \la{scp}
\ee 
in which case $\cos\phi=-f(\beta,v)$.
These are the standard (tree-level) conditions for the existence of
spontaneous CP-violation~\cite{lee}.

Suppose that $\lambda_6,\lambda_7\lsim {\cal O}(g^2)$, as one
would like to have. Then, $m_{12}^2(T)$ must be  
of the order ${\cal O}(g^4T^2)$ to compensate for the terms 
proportional to $\lambda_6 v^2, \lambda_7 v^2$ in \eq\nr{scp}. 
Since $m_{12}^2(T)$ is small, one of the masses
$m_1^2(T),m_2^2(T)$ has to be large, $\sim (gT)^2$, 
in order for the eigenvalues to sum up to \eq\nr{summ}. 
Let us assume\footnote{For completeness, let us note that
in the MSSM it is actually $m_1^2(T)$ which is large, 
but here it is convenient to exchange the fields, to keep 
the discussion as close to \se\ref{sec:diag} as possible.} 
that this role is played by $m_2^2$:
$m_2^2(T) = M^2 + {\cal O}(g^2 T)^2$.
Then, it can be seen from the Higgs part in \eq\nr{H1H2action} that 
\be
\sin\beta \approx -\frac{m_{12}^2 + (1/2) \lambda_6 v^2}{M^2}
\lsim {\cal O}(g^2).
\ee
But this means that the term proportional to $\lambda_7$
in \eq\nr{scp}
can again be neglected, unless $\lambda_7$ is very large, and that 
\eq\nr{scp} becomes 
\be
|f(\beta,v)| \approx \frac{M^2 \sb}{2\lambda_5 \sb v^2}
\approx \frac{M^2}{2 \lambda_5 v^2}. \la{oreql5}
\ee
Thus, we are again lead to the
parametric estimate $\lambda_5\gsim {\cal O }(1)$
in order to have \linebreak $|f(\beta,v)|<1$.
Large numerical values of $v$ (a strong transition)
allow for smaller numerical values of $\lambda_5$.

To summarize, we have found that according to tree-level
considerations, it is not enough to have $\lambda_5>0$ to get spontaneous 
CP-violation, but one should have $\lambda_5\gsim {\cal O}(1)$.
The reason is that with the type of mass parameters and couplings
that appear in the MSSM, the numerator and the denominator in 
\eq\nr{scp} tend to vanish at the same point, when the equations
of motion for $\sin\!\beta$ are taken into account.

\section{The wall profile}
\la{sec:profile}

In the previous section, we saw that even the heavy Higgs
doublet $H$ can have a non-zero (even though small) expectation value 
in the broken phase, and a corresponding non-trivial profile at 
the phase boundary (since $\beta$ typically depends on $v$, 
see \eq\nr{beold}). 
In the usual perturbative language
in terms of the doublets $H_1,H_2$, the possible 
non-trivial profile 
means that the ratio of field values, 
$\tb=v_2/v_1$, can change within the phase boundary.
Since the typical values of $H$ are very small, 
$\<H\>\sim \<h\>\tb \sim g^3T$
(see \eqs\nr{hmagn}, \nr{beold}), $\tb$
is expected to change only by a small amount. 
Nevertheless, in~\cite{hn,non-eq} it is argued 
that the deviations of $\tb$ from a constant value
are important for baryon number generation in the MSSM
(this may not be a necessary requirement, though~\cite{cjk,cm}).
The purpose of this section is to discuss briefly the prospects for 
perturbative and non-perturbative determinations of $\tb$.

Note first that, 
parametrically, the profile of $\tb$, or the expectation
value of $H$, is not computable in perturbation theory. 
This is simply because, as can be seen from \eq\nr{beold}, 
the value of $H$ is determined by that of $h$, 
which in turn is non-perturbative
for $h\sim gT$ (i.e., at least within the phase boundary). 
Numerically, the convergence might of course happen to be reasonably good, 
especially if the transition is strong as is required for baryogenesis.
The wall profile obtained from the 2-loop Landau-gauge effective
potential by solving the classical equations of motion has been 
computed in~\cite{mqs} (see also~\cite{cm,pj}).

Let us then discuss whether the profile of $\tb$
could be determined with lattice simulations. We first
give a gauge-invariant generalization for 
what one may mean with the profile of $\tb$. A static
profile can only exist at the critical temperature. Then 
one may define
\be
\fr12 v_i^2\equiv \< H_i^\dagger H_i\> - 
\left.\< H_i^\dagger H_i\>\right|_\rmi{symm.}, \quad
\tan^2\beta = \frac{v_2^2}{v_1^2}, 
\ee
where $i=1,2$ and the latter expectation value is taken
in the homogeneous ``symmetric'' high-temperature phase. 
This definition is clearly gauge and scale independent. 

Suppose now that we have a lattice geometry such that 
planar interfaces are favoured. There is a zero mode related 
to the location of a planar interface, but this can in principle
be removed. Averaging over the set of configurations 
and over the ($x_1,x_2$)-plane, we get a
value for $\tb(x_3)$. This definition
should not be sensitive to the lattice spacing. However, this
definition is sensitive to the cross-sectional area of the lattice. 

Indeed, denote by $\Delta ({\bf x})$, 
${\bf x}=(x_1,x_2)$, the deviation of the interface
location from the average. Then a configuration 
with $\Delta({\bf x})\neq0$ has the action
\be
S_\rmi{int} \approx \sigma\int d^2x \fr12 [\partial_i \Delta({\bf x})]^2,
\ee
where $\sigma$ is the surface tension.
The fluctuations of the measured projection of the interface location
$\Delta$ are distributed as 
$p(\Delta) = \int d^2x \< \delta (\Delta({\bf x})-\Delta) \>$.
As a consequence, the effective width $l$ of the interface
seen after averaging over the area $A$, is expected to diverge as
\be
l^2 \sim \frac{\int d\Delta \Delta^2 p(\Delta)}{\int d\Delta p(\Delta)} 
\sim \frac{1}{A} \int d^2x \< \Delta({\bf x})^2 \> 
\sim \int d^2k \frac{1}{\sigma k^2}
\sim \frac{1}{\sigma}\ln \frac{A}{\xi^2}, 
\ee
where $\xi$ is of the order of the longest Higgs correlation length.

In conclusion, the profile of the interface can only be studied
by averaging over a finite area.  This can be accomplished either
by restricting the total cross-sectional area, or by averaging
the surface profile only over a finite subarea of the large 
total area (`coarse-graining').
However, we do not regard this as a fundamental problem. 
From the practical point of view, a more serious concern 
is that since the deviations from a 
constant value of $\tb$ are expected to be very small,
they are  most probably not visible from below the 
statistical noise. 

\section{Spontaneous CP-violation}
\la{sec:cp}

In \se\ref{sec:param} we saw that, parametrically, spontaneous
CP-violation is not naturally realized in the MSSM. However, parametric
estimates need not always be reliable due to the actual numerical factors. In
this section, we thus look at the issue from a 
more practical point of view. We will first make some perturbative estimates
and then comment on the possibility of lattice simulations. 

Let us start by formulating the problem as follows. The question 
concerning the existence of spontaneous CP-violation can be factorized
into two parts:
\bi
\item[(1)]
The perturbative computation of the parameters of the action in 
\eq\nr{H1H2action}. This computation is not sensitive to the 
infrared problems of finite temperature field theory and is thus
in principle well convergent. 
\item[(2)]
The determination of the constraint that the parameters 
of the 3d theory have to satisfy in order for there to be
spontaneous CP-violation. As we will discuss, this constraint
is sensitive to the infrared problems of finite temperature
field theory and is ultimately to be determined non-perturbatively.
As a concrete example of the non-perturbative nature of the problem, 
note that the parameter $m_{12}^2(T)$
has a logarithmic scale dependence on the 3d 
renormalization scale parameter $\Lambda$,  
$m_{12}^2(T)\sim (\gamma_1\gamma_{12}+...)T^2/(16\pi^2) 
\ln[\Lambda/(g^2 T)]$, 
and the relevant scale for, say, the statement $m_{12}^2(T)=0$,
cannot be fixed perturbatively. 
\ei

Concerning item (1), we will assume the expressions in 
\eqs\nr{seq1st}--\nr{seqlast} to be precise enough 
for our purposes. If needed, the accuracy can be improved upon
on two aspects, in particular: the zero-temperature 
$\msbar$-scheme parameter $m_{12}^2(\bmu)$ could be 
fixed even more precisely in terms of the physical observables
of the theory by a more accurate vacuum renormalization, 
and the finite temperature corrections
to the parameter $m_{12}^2(T)$ of the effective 3d theory
could be derived up to 2-loop level, to fix its 
scale dependence (see, e.g., \cite{generic}
and the last Ref.\ in~\cite{lo}).

Concerning item (2), we will here work basically at tree-level,
estimating the effects of loop corrections only in the regime 
where they can be reduced to the construction of a further
simplified effective theory. We argue that 
this approximation contains the largest uncertainties
in the present computation,
and we explain in \se\ref{sec:concl} how it can be 
improved upon.

\subsection{Tree-level constraints inside the 3d theory}

The (first) tree-level condition for spontaneous CP-violation
is given in \eq\nr{scp}. To see when it could be satisfied 
in practice, 
note first that for very large fields $v$,
\be
|f(\beta,v)| = \left|\frac{\lambda_6 \ccb+\lambda_7\ssb}
{2\lambda_5\sbb }\right| \gg 1,
\ee
since typically $|\lambda_5| \ll |\lambda_6|,|\lambda_7|$, 
see \eqs\nr{la5}--\nr{aeqlast}. For very small
fields $v\to 0$, on the other hand, 
\be
|f(\beta,v)| = \left| \frac{m_{12}^2}
{\lambda_5\sbb v^2 }\right| \gg 1.
\ee
Moreover, the function 
$(a + b v^2)/(c v^2)$ is a monotonous function 
of $v^2$. Thus there can only be a 
region with $|f(\beta,v)|<1$ provided 
that the numerator of \eq\nr{scp}
is zero at some field values. This requires that 
\be
|m_{12}^2(T)| \sim \{ |\lambda_6|,|\lambda_7| \} v^2,  \la{req}
\ee
and that the sign\footnote{Since we are in this section
considering spontaneous CP-violation, we assume
that $\hat A_t$, $\hat\mu$ are real, but we nevertheless keep the
same notation for them which arises in the complex case.} 
of $m_{12}^2(T)$ is opposite to 
that of $\lambda_6\ccb+\lambda_7\ssb$. 

The second tree-level constraint concerns $\lambda_5$:
\be
\lambda_5>0. \la{lreq}
\ee
The larger $\lambda_5$ is, the larger is the range of values
of $v$ where spontaneous CP-violation can take place.
In \se\ref{eiot} we got a stronger
constraint, $\lambda_5\gsim{\cal O}(g^2)$
(assuming now that $v\sim T$ in \eq\nr{oreql5}, 
instead of $v\sim gT$ as in \se\ref{eiot}).
However, since we will see that the uncertainties in 
the constraint related to $\lambda_5$ are large
(larger than those in the constraint related to $m_{12}^2(T)$
in \eq\nr{req}), we consider here only the most 
relaxed version of the $\lambda_5$-constraint,
the one given in \eq\nr{lreq}. The true
requirement for $\lambda_5$ is eventually to be 
determined numerically. 

Note now that, according to the expressions
for the parameters related to CP-violation 
($m_{12}^2$, $\lambda_5,\lambda_6,\lambda_7$
in \eqs\nr{m12}, \nr{la5}--\nr{aeqlast}),
the inequality in \eq\nr{lreq} can be satisfied. However, 
assuming that $v\lsim T$, 
we see from \eqs\nr{la6}, \nr{la7} 
that \eq\nr{req} goes over into
\be
|m_{12}^2(T)| \lsim \biggl\{ \htttt |\hat A_t^* \hat \mu^*| T^2, 
\gggg \frac{|M_2\mu^*|}{\pi^2}
\biggr\}. \la{reqprac}
\ee
If the expression for $m_{12}^2(T)$ in \eq\nr{m12}
is dominated by contributions from some 
single particle species, then
\eq\nr{reqprac} clearly cannot be satisfied, 
as all the different types of terms in  
\eq\nr{m12} are much larger than needed for
\eq\nr{reqprac}. This holds independent of whether
the dominant effects come from squarks ($\hat A_t, \hat\mu\neq 0$),
gauginos and Higgsinos ($M_2,\mu \neq 0$), or Higgses ($A_t=\mu=0$). 

Thus, the only way to get 
spontaneous CP-violation is to have a cancellation
between the contributions of the different particle species. 
Indeed, one can get $m_{12}^2(T)\sim 0$
at relatively large values of $m_A$, if  
$\mathop{\rm sign}(\hat A_t^*\hat\mu^*) > 0$: 
according to \eqs\nr{m12}, \nr{seq3rd},
\be
m_{12}^2(T) \sim -\fr12 m_A^2 \sbb + 
\fr14 h_t^2 \hat A_t^* \hat \mu^* T^2.
\la{m12new}
\ee
This behaviour is obtained for $m_U^2 \ll (2 \pi T)^2 \ll m_Q^2$.
Numerically, \eq\nr{m12new} could be close to zero if,
for instance, $T\sim T_c\sim 85$ GeV, $\tb \sim 15$, $\hat A_t\sim 0.6$, 
$\hat \mu \sim 0.3$, and $m_A\sim 70$ GeV.\footnote{It is interesting
to note that the negative tree-level term in $m_{12}^2(T)$
can also be compensated 
for by a large stop vev 
$\sim h_t^2 \hat A_t^*\hat \mu^*\langle U^\dagger U\rangle$ 
in the stop breaking phase~\cite{bjls,mssmsim}. However, 
there the Higgses are in the symmetric phase.} 

\subsection{Stop loops inside the 3d theory}
\la{ss:stop}

In the discussion so far, we have completely ignored
loop effects within the effective theory in \eq\nr{H1H2action},
which might change the estimates noticeably (indeed, it is 
only loop effects which induce a first order transition and
a stable phase boundary in the first place). We now proceed
to estimate the loop effects related to the stop
field $U$ 
in the case that it is heavy enough to allow for a perturbative treatment.
By computing its effects on the parameters of the Higgs sector 
and applying then the tree-level estimates in \eqs\nr{req}, \nr{lreq},
we show that the loop effects can be large in the 
regime where $m_{12}^2(T)$ given in \eq\nr{m12new} is small.
On the other hand, if one is clearly outside this regime, then
loop effects cannot trigger spontaneous CP-violation.

To be in the perturbative regime, we assume that the 
stop mass $m_U^2(T)$ is of the parametric order $(g_S T)^2$.
In other words, these estimates do not apply when one is
very close to the ``triple'' point where the direction $U$ 
can get broken (see~\cite{bjls,cqw2,cm}). 

Inside the 3d theory, 1-loop contributions to the masses are of the form 
\be
\int \frac{d^3p}{(2\pi)^3} \frac{1}{p^2+m^2} = -\frac{m}{4 \pi},
\ee
while 1-loop contributions to quartic couplings are of the 
form 
\be
-\int \frac{d^3p}{(2\pi)^3} \frac{1}{[p^2+m_a^2][p^2+m_b^2]}
= -\frac{1}{4\pi (m_a+m_b)}.
\ee
The stop contributions are 
\ba
& & \delta m_{12}^2(T) = -\frac{3}{4\pi} \gamma_{12} T m_U(T)
\sim -h_t^2 \hat A_t^*\hat\mu^* g_S T^2, \la{dm12} \\
& & \delta\lambda_5 = -\frac{3}{16\pi} \frac{\gamma_{12}^2 T}{m_U(T)}
\sim -\frac{h_t^4 (\hat A_t^*\hat \mu^*)^2 T}{g_S T},
\la{stla5} \\
& & 
\delta\lambda_6 = -\frac{3}{8\pi} \frac{\gamma_{12}\gamma_1T}{m_U(T)}, \quad
\delta\lambda_7 = -\frac{3}{8\pi} \frac{\gamma_{12}\gamma_2T}{m_U(T)}.
\la{dla}
\ea

We observe that the contribution to the mass
parameter $m_{12}^2(T)$ is of order ${\cal O}(g^3)T^2$ and is
parametrically small compared with the leading $T^2$-term. 
The contributions to the quartic couplings are, in contrast,
of order ${\cal O}(g^3)$ and thus larger than the corresponding 
tree-level terms in \eqs\nr{la5}--\nr{la7}. 
However, even these corrections are not large enough to make 
\eq\nr{req} naturally satisfied for $v\sim T$, unless the 
zero temperature and $g^2T^2$-terms in $m_{12}^2(T)$
cancel, which was the requirement in \eq\nr{m12new}. 

The negative stop contribution to $\lambda_5$, in fact, 
tends to make spontaneous CP-violation less likely, 
since \eq\nr{lreq} is more difficult to satisfy. 
This effect should be compensated for by the 
gaugino-Higgsino term in \eq\nr{la5}. 

Finally, consider what happens in the limit that the stop 
contributions become non-perturbative, $m_U(T)\sim g_S^2T$. 
In this case, one cannot look at contributions to effective 
parameters, but one may try
to estimate the effects numerically by summing the contributions
to the effective potential. For real parameters
($\hat A_t^* \to \hat A_t$, 
$\hat\mu^* \to \hat\mu$), the 1-loop stop
contribution to the effective potential is of the form 
\be
V \sim 
-\frac{1}{2\pi} \left[ m_U^2(T)
 -\fr12 h_t^2 v_1^2 \muu
+ \fr12 h_t^2 v_2^2 (1-\Att)+ h_t^2 v_1v_2 
\hat A_t\hat\mu \cos\phi 
\right]^{3/2}.
\ee
For large $m_U^2(T)$ and $\mathop{\rm sign}(\hat A_t^*\hat\mu^*) > 0$,
this term is minimized by  $\cos\phi= 1$,
which again corresponds to no CP-violation. However, 
for small $m_U^2(T)$  the term becomes non-analytic and may
induce non-trivial behaviour. At the same time, 
higher order corrections are of the same magnitude
as the 1-loop term.

To summarize the effects of the stop loops: 
$m_{12}^2(T)$ cannot be made small unless it already is so, 
and thus spontaneous CP-violation cannot be triggered far 
away from where it was expected to take place 
according to \eq\nr{m12new}. Moreover, even in this
regime, stop loops in fact seem to make spontaneous 
CP-violation less likely at least as long as they 
allow for a perturbative treatment, 
since they give a large negative
contribution to $\lambda_5$. 

\subsection{Higgs loops inside the 3d theory}

Although the Higgs loops involve effects
from the non-perturbative mass scale $\sim g^2T$, 
it is nevertheless illuminating to inspect what 
kind of contributions they would give to the effective
parameters according to the loop expansion. Note 
that the gauge fields (and $A_0^a, C_0^A$), in contrast, 
do not contribute to the CP-violating couplings at leading order. 

It turns out that the Higgs field contributions 
to the CP-violating parameters are typically 
smaller than the stop contributions. For instance, 
in the limit $m_{12}^2(T)\approx 0$, 
there must be at least one appearance of 
the couplings $\lambda_5$...$\lambda_7$, which are 
of order $h_t^4/(16\pi^2)$ instead of order $h_t^2$ as 
the couplings $\gamma_1,\gamma_2,\gamma_{12}$:
\ba
& & 
\frac{\delta \lambda_5}{T} = 
-\frac{(5/2)\lambda_6^2+2\lambda_1\lambda_5}{8\pi m_1(T)}
-\frac{(5/2)\lambda_7^2+2\lambda_2\lambda_5}{8\pi m_2(T)}
-\frac{6 \lambda_4\lambda_5+4\lambda_3\lambda_5+\lambda_6\lambda_7}
{4\pi[m_1(T)+m_2(T)]}, \\
& & 
\frac{\delta \lambda_6}{T} = 
-\frac{12 \lambda_1\lambda_6}{8\pi m_1(T)}
-\frac{3\lambda_3\lambda_7+2\lambda_4\lambda_7+2\lambda_5\lambda_7^*}
 {8\pi m_2(T)}
-\frac{3\lambda_3\lambda_6 + 4\lambda_4\lambda_6+10 \lambda_5\lambda_6^*}
{4\pi[m_1(T)+m_2(T)]}, \\
& & 
\frac{\delta \lambda_7}{T} = 
-\frac{3 \lambda_3\lambda_6+2\lambda_4\lambda_6+2\lambda_5\lambda_6^*}
{8\pi m_1(T)}
-\frac{12 \lambda_2\lambda_7}{8\pi m_2(T)}
-\frac{3\lambda_3\lambda_7+4\lambda_4\lambda_7+10\lambda_5\lambda_7^*}
{4\pi[m_1(T)+m_2(T)]}. \hspace*{1cm}
\ea
The terms where there appears only the light Higgs with mass $\sim g^2T$
are seen to contain non-perturbative contributions which are parametrically 
as large as the original couplings. There are also other types of
contributions of similar magnitude: 
expanding formally in $m_{12}^2(T)\neq 0$, 
one gets for example terms of the type 
\be
\frac{\delta \lambda_6}{T} \sim 
\frac{\lambda^2 m_{12}^2(T)}{4\pi m_i(T)[m_1(T)+m_2(T)]^2},  
\ee
where $\lambda\sim \{\lambda_1,...,\lambda_4\}$, 
$i=1,2$. 
All these contributions involve expansion parameters 
$\sim \{\lambda_1,...,\lambda_4\}T/(g^2 T)\sim 1$, and
are thus not perturbatively computable. 

\subsection{Numerical values}

Let us finally look at 
the numerical requirements for the parameter 
values after taking into account the dimensionally 
reduced effective couplings in \se\ref{summary}, 
together with the stop contributions
in the regime where they are still 
assumed to be perturbative, \se\ref{ss:stop}. 
To compensate
for the negative stop contribution to $\lambda_5$ in 
\eq\nr{stla5} by a positive 
gaugino-Higgsino contribution
in \eq\nr{seql5}, one needs that 
\bi
\item
$m_Q$ is large, 
because then the gaugino-Higgsino term suppressed
only by temperature is relatively more significant; 
\item
$M_2$ is large in order to increase the gaugino-Higgsino
contribution, but $\hat A_t$ is not very large, to 
decrease the stop contribution.
\ei
On the other hand, to allow for 
relatively large values of $m_A$ at the point
where $m_{12}^2(T)$ in \eq\nr{seq3rd} is small
(for fixed $T\sim T_c$), it is preferable to have 
$\mathop{\rm sign}(M_2 \mu^*) < 0$ and $|\mu|$ large
(this may help also in making $\lambda_6,\lambda_7$
positive, so that $m_{12}^2(T)$ may remain slightly 
negative).

Using the expressions in \se\ref{summary} and
those in \eqs\nr{dm12}, \nr{stla5}, we estimate that 
numerically the best region for these requirements is
\ba
& & 
m_Q \gsim 800\mbox{ GeV},\quad
M_2\gsim 70\mbox{ GeV}, \quad    
|\mu|\gsim 50\mbox{ GeV}, \quad
|\hat A_t| \lsim 0.2,  \la{regime}
\ea
where $\hat A_t$ is as defined in \eq\nr{Atmudef}.
In these estimates, we have assumed that $T\sim T_c\sim 85$ GeV, 
as can be observed by equating \eq\nr{seq2nd} with zero
in the limit that $\tb$ is large 
(for numerical estimates of $T_c$ 
based on the 2-loop potential and lattice see, 
e.g., \cite{bjls,lo,mssmsim}). 
Nevertheless, even in the regime of \eq\nr{regime} 
there remains the constraint
\be
m_A^2 \sbb \lsim 400 \mbox{ GeV}^2. \la{mAreg}
\ee
Hence one needs, say, $m_A\lsim 65$ GeV 
and $\tb\gsim 20$. Clearly
this parameter region is at most barely 
consistent with what is required for the 
strength of the transition in order to preserve the baryon 
asymmetry generated, and with experimental constraints, 
which both provide a lower bound for $m_A$~\cite{cm}. 

\subsection{Prospects for lattice simulations}

In the case of explicit CP-violation, some of the 
parameters appearing in the action have complex phases and the 
expectation values of CP-violating operators such as $O_4=O_\rmi{CP}$
in \eq\nr{ops} are proportional to these. 
Due to the experimental constraints
on the complex phases which are thought to be 
relatively strong, the magnitude of explicit CP-violation should
be relatively small for realistic parameter values.
In this case, a lattice study of the CP-violating
effects appears rather difficult from the practical point of view, 
for the same reasons as discussed in \se\ref{sec:profile} for $\tb$. 
Perhaps the best that could be done is to solve the 
classical equations of motion for the CP-violating wall profile
using the 2-loop effective potential.  Even though the effects
may be small, this case is of
interest since it has been argued
that explicit CP-violation 
might suffice for baryogenesis~\cite{non-eq,cjk}. 
Note also that large explicit effects
may not be excluded~\cite{bgk}, in which 
case also a lattice study is feasible.

In the case of spontaneous CP-violation, on the other hand, the 
parameters can be real and yet the effects can be large and 
well detectable: the distribution of $O_\rmi{CP}$ 
should have two peaks at opposite values of $O_\rmi{CP}$
(perhaps only within the phase boundary), 
which gives an unambiguous signal. 
The role of explicit CP-violation is merely to choose one of
the peaks as the physically preferred one.

\section{Conclusions}
\la{sec:concl}

In this paper, we have studied the general prospects for 
spontaneous CP-violation around the electroweak phase
transition in the MSSM. Parametrically, spontaneous 
CP-violation seems not to be easily realized. Nevertheless, 
we have shown that numerically there is 
a small regime, given in \eqs\nr{regime}, \nr{mAreg}, which 
could be favourable for spontaneous CP-violation.
We have shown that in this regime, the IR-sensitive loop contributions 
related to the stops are large and even non-perturbative
in the light stop regime. The Higgs field contributions
are non-perturbative, as well, but are numerically
smaller than the stop contributions. However, it does not
seem obvious that the large effects would naturally increase
the likelihood of spontaneous CP-violation. It is 
rather that one needs delicate cooperation from 
different particle species to get an effect. Moreover, 
there is the concern that the regime found for $m_A$ within 
the present approximations, is almost certainly excluded due 
to experimental constraints and the requirement for 
a strong 1st order phase transition~\cite{cm}. 

It is conceivable that in many extensions of the MSSM, 
such as the NMSSM~\cite{nmssm} 
and various two Higgs doublet models, the parameter region allowing 
for spontaneous CP-violation would be larger than in the MSSM.
Moreover, even if there is no spontaneous CP-violation, it could
be that in the case of explicit CP-violation, the 
non-perturbative dynamics of the 
3d theory affects strongly the explicit effects
(especially if the explicit effects are large, 
as proposed in~\cite{bgk}). It has also
been argued that there could be CP-violating solitons
in the broken phase of the theory, even though 
there is no spontaneous CP-violation~\cite{rt}. Thus, in spite of the
fact that the parameter space for spontaneous CP-violation 
in the MSSM seems to be quite small, we believe that a more
precise study of these questions is motivated.
We expect relatively large uncertainties in the present results.  

The present results were based on deriving an effective 
action for the IR degrees of freedom of the theory, and 
studying then this effective action at the tree-level.
The first part thereof, 
the derivation of the effective action, is a purely 
perturbative computation, but even its accuracy 
can in principle be improved upon. In particular, the $\msbar$ 
parameter $m_{12}^2(\bmu)$ could be related more precisely to 
physical observables such as $m_A$, and the parameter 
$m_{12}^2(T)$ of the effective 3d theory could be 
derived with 2-loop accuracy. Nevertheless, we expect
that it is the non-perturbative
IR-properties of the theory that are responsible 
for the largest uncertainties.

The first step of a more precise IR-study 
would be to compute the full 1-loop and 2-loop effective 
potentials within the effective theory, and then to
solve the complete classical equations of motion for the
wall profile consisting of  $v_1,v_2,\phi$,
to see when~$\phi$ can deviate from zero. Solving 
the classical equations of motion is in principle
straightforward but can lead, in practice, to somewhat 
difficult instabilities and slow convergence;
a working algorithm has been recently proposed in~\cite{pj}.

At the second step, the theory in \eq\nr{H1H2action} could 
be studied with lattice simulations. Concerning
the lattice study, it should be realized that the 
wall profile can only be defined with a finite 
cross-sectional area, see \se\ref{sec:profile}. 
However, we do not consider this to be a fundamental problem
(the area dependence is only logarithmic). 
It should be noted that 
a lattice study of spontaneous CP-violation is,
from a practical point of view, more promising than 
for instance a study of the non-trivial profile of $\tb$, 
since the effect can be observed by finding a non-trivial
(two-peak) distribution of a single operator, such as
$O_\rmi{CP}$ in \eq\nr{ops}. The breaking of global 
discrete symmetries in 3d gauge theories
has previously been studied numerically in~\cite{adjoint},
for the SU(3)+adjoint Higgs model.

Finally, it is interesting to note that the leading parity
violating effects are very difficult to study with lattice
simulations, since they are induced by higher-dimensional
operators in the 3d theory 
which are complex in Euclidian space~\cite{pviol}.
The C-violating (and thus CP-violating) effects discussed 
in this paper are, in contrast, contained in a real 
super-renormalizable Euclidian action, and there 
should be no fundamental problems in the simulations. 

\section*{Acknowledgements}

M.L. thanks M. Losada, A. Pilaftsis and
A. Riotto for useful discussions. This work 
was partly supported by the TMR network 
{\em Finite Temperature Phase Transitions in Particle
Physics}, EU contract no.\ FMRX-CT97-0122.

\appendix
\renewcommand{\thesection}{Appendix~~\Alph{section}}
\renewcommand{\thesubsection}{\Alph{section}.\arabic{subsection}}
\renewcommand{\theequation}{\Alph{section}.\arabic{equation}}

\section{}

In this Appendix, we derive the 1-loop expressions for the 
parameters in \eq\nr{H1H2action}. CP-violation is assumed
to appear in the original theory (at some renormalization 
scale $\bmu_0$) only in the trilinear couplings $A_t,\mu$,
defined in terms of the Euclidian 4d Lagrangian as 
in \eq\nr{CPparams}.
Thus, in particular, the Higgs sector mass parameter $m_{12}^2$
and the gaugino mass parameter $M_2$ are assumed to have been
tuned real at $\bmu_0$, employing field redefinitions.
Of the parameters in \eq\nr{CPparams}, 
$\mu$ is taken to be equal to the supersymmetric mass 
parameter which appears also in the Higgsino mass matrix, 
although in principle there could be 
an arbitrary soft component in the trilinear term
to which $\mu$ contributes~\cite{rosiek}. The
soft parameter $A_t$, in contrast, only appears in the 
interaction in \eq\nr{CPparams}.

We shall work in the limit $m_i^2\ll (2\pi T)^2 \ll m_Q^2$, 
$A_t^2\ll m_Q^2$, where $m_i^2$ signals all the mass parameters 
of the theory other than $m_Q^2$: $m_i^2 = 
\{m_1^2,m_2^2,m_{12}^2,m_U^2,|\mu|^2, M_2^2,M_3^2\}$. Note that 
the trilinear coupling $A_t$ need not, in principle, be smaller
than $2 \pi T$.

In this limit, the construction of the effective theory can be
carried out in two steps. First, in \se\ref{mQ},
we integrate out the left-handed
squark field $Q$ at zero temperature. 
Second, we go to finite temperature and integrate out
the non-zero Matsubara modes of the remaining bosons (\se\ref{bosons}),
as well as all the fermions (\se\ref{fermions}). For the
sake of brevity, we discuss the field redefinitions related 
to all these steps at once in \se\ref{redef}. In \se\ref{gauge}
we discuss issues which are somewhat less central to
the present problem, namely the values of the gauge 
couplings  and the removal of the zero Matsubara 
modes of the gauge fields $A_0,C_0$.
Finally, for completeness, we discuss the most important effects
related to vacuum renormalization in \se\ref{vacuum}. 
All the results are collected together and 
simplified final expressions for the parameters
appearing in \eq\nr{H1H2action}
are given in \se\ref{summary}.

It should be mentioned that many of the expressions given 
here have also been derived in other places. In particular, 
the dimensional reduction steps (but for $m_Q^2\ll (2 \pi T)^2$
and mostly only for the Higgs sector) have
been carried out in~\cite{ml}--\cite{lo}
(see also \cite{joa}). In addition, 
expressions for the quadratic and quartic Higgs couplings 
involving CP-violation have been derived,
in various limits, in~\cite{emq,cpr,fkot}.

The notation for the parameters of the final effective action, 
to be used below, is
given in \eq\nr{H1H2action}. In addition 
to the effective parameters, the individual reduction steps
also give contributions to the wave functions, which we 
denote by
\ba
\delta {\cal L} & = & 
(D_i^w H_1)^\dagger(D_i^w H_1) \delta Z_{1} + 
(D_i^w H_2)^\dagger(D_i^w H_2) \delta Z_{2} + 
(D_i^s U)^\dagger (D_i^s U) \delta Z_U \nn
& + &  \Bigl[ (D_i^w H_1)^\dagger(D_i^w \tilde H_2) \delta Z_{12} 
+\Hc\Bigr]. 
\la{wavefunction}
\ea
We do not write down the $\msbar$ scheme 
divergences $1/\epsilon$ in any of the expressions;
these can be restored by $\ln\bmu^2\to 1/\epsilon + \ln\bmu^2$.
In the loop corrections, we always neglect
terms proportional to the U(1) gauge coupling $g'^2$.

\subsection{Integrating out $Q$ at $T=0$}
\la{mQ}

Let us consider the case that the mass parameter related to the 
left-handed stops is much larger than the temperature, 
$m_Q \gg 2 \pi T$. Then, the field
$Q$ can be integrated out from the zero temperature
theory. Note that this case is different from the 
consideration in~\cite{ceqw}, where all the squarks
were integrated out with degenerate mass parameters.
Evaluating some typical finite temperature integrals 
numerically, we expect the results to be well
reliable when, say, $m_Q\gsim 0.6$ TeV
($T_c \lsim 100$ GeV).

The left-handed stop field $Q$ contributes to the effective
action of the two Higgses $H_1,H_2$ and the right-handed stop $U$
already at tree-level, due to graphs of the type
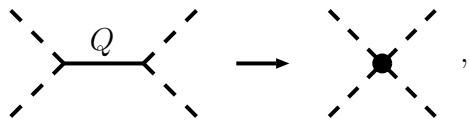
\begin{center}
\begin{picture}(180,60)(0,0)

\SetWidth{1.5}
\DashLine(10,10)(30,30){5}
\DashLine(10,50)(30,30){5}
\Line(30,30)(60,30)
\DashLine(60,30)(80,50){5}
\DashLine(60,30)(80,10){5}
\Text(40,38)[l]{$Q$}

\LongArrow(95,30)(115,30)

\DashLine(130,10)(170,50){5}
\DashLine(130,50)(170,10){5}
\GCirc(150,30){3}{0}
\Text(180,30)[l]{,}

\end{picture}
\end{center}
where the dashed line denotes the light degrees of freedom, 
the solid line the heavy field $Q$, and the filled circle
an effective vertex.

Using the notation $\hat A_t,\hat\mu$ defined in \eq\nr{Atmudef}, 
the tree-level couplings of the action in \eq\nr{H1H2action}, 
after integrating out $Q$, are 
\ba
& & \lambda_U = \frac{g_S^2}{6}, \quad
\gamma_1=-h_t^2\hat\mu^2, \quad
\gamma_2=h_t^2 (1-\hat A_t^2), \quad
\gamma_{12}=h_t^2 \hat A_t^*\hat \mu^*, \nn
& & \lambda_1=\lambda_2=\frac{g^2+g'^2}{8},\quad
\lambda_3=\frac{g^2-g'^2}{4},\quad
\lambda_4=-\frac{g^2}{2}, \nn
& & \lambda_5=\lambda_6=\lambda_7=0,   \la{treelevel}
\ea
where $h_t$ is the top Yukawa coupling, and 
$g_S,g,g'$ are the strong, weak and U(1) gauge couplings. 
There are no contributions to the wave functions in 
\eq\nr{wavefunction} at tree-level.

At 1-loop level, the following types of contributions arise
(we expand in $m^2/m_Q^2$, $q^2/m_Q^2$):

\vspace*{0.4cm}

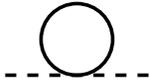
\parbox[c]{5.4cm}{
\begin{picture}(140,40)(0,0)

\SetWidth{1.5}
\SetScale{0.9}
\DashLine(0,5)(60,5){5}
\CArc(30,20)(15,0,360)

\end{picture}}
\parbox[c]{9.5cm}{
\ba
& & \pint \frac{1}{p^2+m_Q^2}
=-\frac{m_Q^2}{16\pi^2}
\biggl(\ln\frac{\bmu^2}{m_Q^2}+1\biggr), \hspace*{2.2cm} \la{mQ1st}
\ea}



\parbox[c]{5.4cm}{
\begin{picture}(140,40)(0,0)

\SetWidth{1.5}
\SetScale{0.9}
\DashLine(0,20)(20,20){5}
\DashLine(50,20)(70,20){5}
\CArc(35,20)(15,180,360)
\DashCArc(35,20)(15,0,180){5}
\Text(2,28)[l]{$q$}
\Text(45,35)[l]{$m$}

\Line(82,20)(88,20)

\DashLine(100,20)(140,20){5}
\DashCArc(120,30)(10,0,360){5}
\GCirc(120,20){3}{0}

\end{picture}}
\parbox[c]{9.5cm}{
\ba
& & \pint \frac{1}{[p^2+m_Q^2][(p+q)^2+m^2]} -
\frac{1}{m_Q^2}\pint \frac{1}{p^2+m^2}
\nn
& & 
=\pf \biggl[
\biggl(1+\frac{m^2}{m_Q^2}\biggr)
\biggl(\ln\frac{\bmu^2}{m_Q^2}+1\biggr)-\fr12\frac{q^2}{m_Q^2}
\biggr], 
\ea}



\parbox[c]{5.4cm}{
\begin{picture}(140,40)(0,0)

\SetWidth{1.5}
\SetScale{0.9}
\DashLine(0,0)(20,20){5}
\DashLine(0,40)(20,20){5}
\DashLine(50,20)(70,0){5}
\DashLine(50,20)(70,40){5}
\CArc(35,20)(15,0,360)

\end{picture}}
\parbox[c]{9.5cm}{
\ba
& & 
\pint \frac{1}{(p^2+m_Q^2)^2}
=\pf \ln\frac{\bmu^2}{m_Q^2}, \hspace*{3cm}
\ea} 



\parbox[c]{5.4cm}{
\begin{picture}(150,40)(0,0)

\SetWidth{1.5}
\SetScale{0.9}
\DashLine(0,0)(20,0){5}
\DashLine(0,40)(20,40){5}
\DashLine(20,0)(60,40){5}
\DashLine(20,40)(60,0){5}
\Line(20,0)(20,40)

\Line(67,20)(73,20)

\DashLine(80,0)(100,20){5}
\DashLine(80,40)(100,20){5}
\DashLine(130,20)(150,40){5}
\DashLine(130,20)(150,0){5}
\DashCArc(115,20)(15,0,360){5}
\GCirc(100,20){3}{0}

\end{picture}}
\parbox[c]{9.5cm}{
\ba
& & 
\pint \frac{1}{(p^2+m_Q^2)(p^2+m^2)^2} -
\frac{1}{m_Q^2}\pint \frac{1}{(p^2+m^2)^2}
\nn
& & 
=- \frac{1}{m_Q^2}\pf
\biggl(\ln\frac{\bmu^2}{m_Q^2}+1\biggr), \hspace*{2.5cm}
\ea} 



\parbox[c]{5.4cm}{
\begin{picture}(140,40)(0,0)

\SetWidth{1.5}
\SetScale{0.9}
\DashLine(0,0)(20,0){5}
\DashLine(0,40)(20,40){5}
\Line(20,0)(40,20)
\DashLine(40,20)(60,40){5}
\Line(20,40)(40,20)
\DashLine(40,20)(60,0){5}
\DashLine(20,0)(20,40){5}

\Line(72,20)(78,20)

\DashLine(90,0)(130,40){5}
\DashLine(90,40)(130,0){5}
\DashCArc(98,20)(8,70,290){5}
\GCirc(110,20){3}{0}

\end{picture}}
\parbox[c]{9.5cm}{
\ba
& & \pint \frac{1}{(p^2+m_Q^2)^2 (p^2+m^2)} -
\frac{1}{m_Q^4}\pint \frac{1}{p^2+m^2}
\nn
& & 
=\frac{1}{m_Q^2} \pf, \la{six} \hspace*{5.0cm}
\ea}



\parbox[c]{5.4cm}{
\begin{picture}(150,40)(0,0)

\SetWidth{1.5}
\SetScale{0.9}
\DashLine(0,0)(15,5){5}
\DashLine(0,40)(15,35){5}
\DashLine(15,5)(45,5){5}
\DashLine(15,35)(45,35){5}
\DashLine(45,5)(60,0){5}
\DashLine(45,35)(60,40){5}
\Line(15,5)(15,35)
\Line(45,5)(45,35)

\Line(67,20)(73,20)

\DashLine(80,0)(100,20){5}
\DashLine(80,40)(100,20){5}
\DashLine(130,20)(150,40){5}
\DashLine(130,20)(150,0){5}
\DashCArc(115,20)(15,0,360){5}
\GCirc(100,20){3}{0}
\GCirc(130,20){3}{0}

\end{picture}}
\parbox[c]{9.5cm}{
\ba
& & 
\pint \frac{1}{(p^2+m_Q^2)^2(p^2+m^2)^2} -
\frac{1}{m_Q^4}\pint \frac{1}{(p^2+m^2)^2}
\nn
& & 
=-\frac{1}{m_Q^4} \pf 
\biggl(\ln\frac{\bmu^2}{m_Q^2}+2\biggr), \la{mQlast} \hspace*{3.5cm}
\ea}

\vspace*{0.4cm}

\noindent
where $dp\equiv d^{4-2\epsilon} p/(2\pi)^{4-2\epsilon}$.
Here the subtracted parts are contributions within the effective
theory: after the subtractions the results
are IR-safe as they should be. The effective six-point vertex
in \eq\nr{six} is shown just for illustration;
it gives a contribution suppressed by the relative
amount $m^2/m_Q^2$ and is not kept in the 
actual theory in \eq\nr{H1H2action}.
 
A 1-loop computation
of all the graphs of the types in \eqs\nr{mQ1st}--\nr{mQlast}
results in the following contributions to the
effective parameters in \eqs\nr{aeq1st}--\nr{aeqlast}
and to the wave functions in \eq\nr{wavefunction}:
\ba
& & 
\delta Z_1=\fr32 \htt |\hat\mu|^2,\quad 
\delta Z_2=\fr32 \htt |\hat A_t|^2,\\
& & 
\delta Z_{12}=-\fr32 \htt \hat A_t^* \hat\mu^*,\quad 
\delta Z_U= \htt(|\hat A_t|^2+|\hat\mu|^2), \\
& & 
\delta m_1^2(T) = -3 \htt \lnmQQ (|\mu|^2 + |\hat\mu|^2 m_U^2),
\la{mQmm1} \\
& & 
\delta m_2^2(T) = 
-3 \htt \lnmQQ (m_Q^2+|A_t|^2+|\hat A_t|^2 m_U^2), \la{mQmm2}\\
& & 
\delta m_{12}^2(T) = 
3 \htt \lnmQQ (A_t^*\mu^* + \hat A_t^*\hat \mu^* m_U^2), \\
\la{mQm12}
& &
\delta m_U^2(T) = -2 \htt\lnmQQ \Bigl(m_Q^2+|A_t|^2+|\mu|^2 \nn
& & \hspace*{1cm} +
|\hat A_t|^2 m_2^2 + |\hat \mu|^2 m_1^2-
(\hat A_t^*\hat\mu^*+\hat A_t\hat\mu) m_{12}^2\Bigr), \la{mQmU} \\
& & 
\delta \lambda_U = \pf\biggl[
(-h_t^4+\fr23 h_t^2 g_S^2-\fr16 g_S^4)\lnmQ 
-\frac{1}{12} g_S^4 \ln\frac{\bmu^2}{m_D^2}
+2 h_t^4\biggl(|\hat\mu|^2  \nn
& & \hspace*{1cm} -|\hat A_t|^2\lnmQ \biggr)
-\fr23 h_t^2 g_S^2 (|\hat A_t|^2+|\hat\mu|^2)+
h_t^4 (\Att+\muu)^2 \lnmQQQ \biggr],\hspace*{1.1cm} \\
& & 
\delta \gamma_1 = \htt \muu \biggl[
h_t^2 -\biggl(\fr43 g_S^2+\fr34 g^2\biggr)\lnmQQ \nn
& & \hspace*{1cm} 
+ h_t^2 (\Att+\muu) \lnmQQQ \biggr], \\
& & 
\delta \gamma_2 = \htt \biggl[
-h_t^2 \lnmQ+h_t^2 (2\Att+\muu)-
\Att \biggl(\fr43g_S^2+\fr34 g^2\biggr)\lnmQQ \nn
& & \hspace*{1cm}  + h_t^2 \Att(\Att+\muu)\lnmQQQ
\biggr], \\
& & 
\delta \gamma_{12} = 
\htt \hat A_t^*\hat\mu^* \biggl[
-h_t^2 +\biggl(\fr43 g_S^2-\fr34 g^2 \biggr)\lnmQQ  \nn
& & \hspace*{1cm} 
- h_t^2 (\Att+\muu) \lnmQQQ \biggr], \\
& & 
\delta \lambda_1 = \pf\biggl\{
h_t^2\muu \biggl[
\fr34 g^2 + \fr32 h_t^2 \muu \lnmQQQ
\biggr] - \frac{3}{16}g^4\lnmQ\biggr\}, \\
& & 
\delta \lambda_2 = \pf\biggl\{ h_t^2 \biggl[
\biggl(-\fr32 h_t^2 -3 h_t^2 \Att  + \fr34 g^2
\biggr)\lnmQ-
\fr34 g^2 \Att \nn
& & \hspace*{1cm} 
+\fr32 h_t^2 |\hat A_t|^4 \lnmQQQ
\biggr] - \frac{3}{16}g^4\lnmQ\biggr\}, \\
& & 
\delta \lambda_3 = \pf \biggl\{ h_t^2 \biggl[
\fr34 g^2 \lnmQ -3 h_t^2 \muu \lnmQQ+\fr34 g^2 (\muu-\Att) \nn
& & \hspace*{1cm} 
+  3 h_t^2 \Att \muu \lnmQQQ
\biggr]  - \frac{3}{8}g^4\lnmQ\biggr\}, \\
& & 
\delta \lambda_4 = \pf\biggl\{ h_t^2 \biggl[
-\fr32 g^2 \lnmQ +3 h_t^2 \muu +\fr32 g^2 (\Att-\muu) \nn
& & \hspace*{1cm}  + 
3 h_t^2 \Att \muu \lnmQQQ
\biggr] + \frac{3}{4}g^4\lnmQ\biggr\}, \\
& & 
\delta \lambda_5 = \fr32 \htttt (\hat A_t^*\hat\mu^*)^2\lnmQQQ, \\
& & 
\delta \lambda_6 = -\htt \hat A_t^*\hat\mu^*\biggl[
\fr34 g^2  +
3 h_t^2 \muu \lnmQQQ
\biggr], \\
& & 
\delta \lambda_7 = \htt \hat A_t^*\hat\mu^* \biggl[
3 h_t^2 \lnmQ + \fr34 g^2 -3 h_t^2 \Att \lnmQQQ
\biggr].
\ea

\subsection{Integrating out bosonic non-zero Matsubara modes}
\la{bosons}

The bosonic non-zero Matsubara modes are integrated out in 
the theory defined by the tree-level couplings in \eq\nr{treelevel}.
Thus, there are loop contributions involving Higgses, right-handed
stops and the gauge fields. 

The types of graphs and integrals appearing have been discussed
in~\cite{generic} (see also~\cite{ml}--\cite{lo}). 
For completeness, let us briefly review the results
here, as well. We use the following notation:
\ba
\muT & = &  4\pi e^{-\gamma} T \approx 7.0555T, \\
\Tint{p_b}' \frac{1}{(p^2)^3} &  = &  
\frac{\zeta(3)}{128\pi^4T^2} \equiv \frac{B_6}{16\pi^2}, \quad
B_6 = \frac{\zeta(3)}{2}\frac{1}{(2\pi T)^2}, \\
\Tint{p_b}' \frac{1}{(p^2)^4} &  = &  
\frac{\zeta(5)}{1024\pi^6T^4} \equiv \frac{B_8}{16\pi^2}, \quad
B_8 = \frac{\zeta(5)}{4}\frac{1}{(2\pi T)^4},
\ea
where $p_b$ denotes the bosonic Matsubara momenta and
a prime means that the zero Matsubara mode is omitted.
Then the basic contributions are of the types

\vspace*{0.4cm}

\parbox[c]{3.0cm}{
\begin{picture}(60,40)(0,0)

\SetWidth{1.5}
\SetScale{0.9}
\DashLine(0,10)(60,10){5}
\DashCArc(30,25)(15,0,360){5}
\Text(50,35)[l]{$n\neq 0$}

\end{picture}}
\parbox[c]{11.9cm}{
\ba
& & \Tint{p_b}' \frac{1}{p^2+m^2} = 
\frac{T^2}{12} - \frac{m^2}{16\pi^2} \Lb, \hspace*{3.6cm}
\ea}



\parbox[c]{3.0cm}{
\begin{picture}(60,40)(0,0)

\SetWidth{1.5}
\SetScale{0.9}
\DashLine(0,10)(60,10){5}
\PhotonArc(30,25)(15,0,360){1.5}{16}
\Text(50,35)[l]{$n\neq 0$}

\end{picture}}
\parbox[c]{11.9cm}{
\ba
& & 
\Tint{p_b}' \biggl(
\frac{\delta_{\mu\mu}}{p^2}-\frac{p_\mu p_\mu}{(p^2)^2}\biggr) = 
\frac{T^2}{4}, \hspace*{4.95cm}
\ea}



\parbox[c]{3.0cm}{
\begin{picture}(70,40)(0,0)

\SetWidth{1.5}
\SetScale{0.9}
\DashLine(0,20)(20,20){5}
\DashLine(50,20)(70,20){5}
\PhotonArc(35,20)(15,180,360){1.5}{8}
\DashCArc(35,20)(15,0,180){5}
\Text(2,28)[l]{$q$}
\Text(50,35)[l]{$n\neq 0$}

\end{picture}}
\parbox[c]{11.9cm}{
\ba
& & 
\Tint{p_b}' \frac{(2q_\mu+p_\mu)(2q_\nu+p_\nu)}{[p^2][(p+q)^2+m^2]}
\biggl(\delta_{\mu\nu}-\frac{p_\mu p_\nu}{p^2}\biggr) 
\nn
& & 
=\frac{3 q^2}{16\pi^2} \biggl(
\Lb-\frac{10}{9} m^2 B_6 \biggr), \hspace*{4.3cm}
\ea}



\parbox[c]{3.0cm}{
\begin{picture}(70,40)(0,0)

\SetWidth{1.5}
\SetScale{0.9}
\DashLine(0,0)(20,20){5}
\DashLine(0,40)(20,20){5}
\DashLine(50,20)(70,0){5}
\DashLine(50,20)(70,40){5}
\DashCArc(35,20)(15,0,360){5}
\Text(20,40)[l]{$n\neq 0$}

\end{picture}}
\parbox[c]{11.9cm}{
\ba
& & 
\Tint{p_b}'\frac{1}{(p^2+m_1^2)(p^2+m_2^2)}
\\
& & 
=\pf \biggl[
\Lb-(m_1^2+m_2^2) B_6 +(m_1^4+m_1^2m_2^2+m_2^4) B_8 \biggr], \nonumber
\ea}


\parbox[c]{3.0cm}{
\begin{picture}(70,40)(0,0)

\SetWidth{1.5}
\SetScale{0.9}
\DashLine(0,0)(20,20){5}
\DashLine(0,40)(20,20){5}
\DashLine(50,20)(70,0){5}
\DashLine(50,20)(70,40){5}
\PhotonArc(35,20)(15,0,360){1.5}{16}
\Text(20,40)[l]{$n\neq 0$}

\end{picture}}
\parbox[c]{11.9cm}{
\ba
& & 
\Tint{p_b}'\frac{\Bigl(\delta_{\mu\nu}-\frac{p_\mu p_\nu}{p^2}\Bigr)
\Bigl(\delta_{\mu\nu}-\frac{p_\mu p_\nu}{p^2}\Bigr)}{(p^2)^2}
=\pf \biggl[
3 \Lb- 2\biggr]. \hspace*{1cm}
\ea}

\vspace*{0.4cm}

\noindent
Here the external momentum is $q=(0,{\bf q})$.
Note that in contrast to \se\ref{mQ}, there are no
subtractions to be made at 1-loop level: due to momentum
conservation, {\em all} the internal lines have a non-zero
Matsubara mode and are thus heavy.

In most places, 
we keep only the dominant terms of the different
qualitative types
in the high temperature expansion in 
$m_1^2/(2\pi T)^2,m_2^2/(2\pi T)^2,m_{12}^2/(2\pi T)^2,m_U^2/(2\pi T)^2$.
Note that to obtain these expressions, it is convenient to write 
the Higgs field propagators as 
\be
\< H_m^{i*}H_n^j\> = \delta_{mn}\delta^{ij}
\biggl(\frac{1}{p^2} - \frac{m_n^2}{(p^2)^2}\biggr), \quad
\< H_m^{i*}H_n^{j*} \> = \epsilon_{mn}\epsilon^{ji} \frac{m_{12}^2}{(p^2)^2}.
\ee
We then obtain that
\ba
& & 
\delta Z_1=-\fr94\gpf\Lb,\quad
\delta Z_2=-\fr94\gpf\Lb,\\
& & 
\delta Z_{12}=\fr52 \gpf m_{12}^2 B_6, \quad
\delta Z_U=-4 \frac{g_S^2}{16\pi^2} \Lb, \\
& & 
\delta m_1^2= 
\frac{3}{16}g^2T^2 - 3 h_t^2\muu \biggl(
\frac{T^2}{12}-\frac{m_U^2}{16\pi^2}\Lb\biggr)+
\fr34 g^2 \biggl(
\frac{T^2}{12}-\frac{m_1^2}{16\pi^2}\Lb\biggr), \\
& & 
\delta m_2^2= 
\frac{3}{16}g^2T^2 + 3 h_t^2(1-\Att) \biggl(
\frac{T^2}{12}-\frac{m_U^2}{16\pi^2}\Lb\biggr) \nn
& & \hspace*{1cm} +
\fr34 g^2 \biggl(
\frac{T^2}{12}-\frac{m_2^2}{16\pi^2}\Lb\biggr), 
\la{bosonmm2} \\
& & 
\delta m_{12}^2 = 3 h_t^2 \hat A_t^*\hat \mu^* \biggl(
\frac{T^2}{12}-\frac{m_U^2}{16\pi^2}\Lb\biggr)+
\fr34 g^2 \frac{m_{12}^2}{16\pi^2}\Lb, \la{bosonm12} \\
& & 
\delta m_U^2 = \fr13 g_S^2 T^2 +\fr43 g_S^2 \biggl(
\frac{T^2}{12}-\frac{m_U^2}{16\pi^2}\Lb\biggr)+
\fr16 h_t^2 (1-\Att-\muu) T^2 \nn 
& & \hspace*{1cm} + 
2 \htt 
\Bigl[\muu m_1^2-(1-\Att) m_2^2 - (\hat A_t^*\hat \mu^* + 
\hat A_t \hat\mu) m_{12}^2 \Bigr]
\Lb, \\
& & 
\delta \lambda_U = 
\pf\biggl\{
-h_t^4 \biggl[
\biggl(|\hat\mu|^4+(1-\Att)^2+2 \Att\muu\biggr)\Lb  - 
2 m_{12}^2 B_6 (1 \nn
& & \hspace*{1cm} -\Att-\muu) (\hat A_t^*\hat \mu^*+\hat A_t\hat\mu)
\biggr] - g_S^4 \biggl[ \frac{13}{12}
\biggl(\Lb-\fr23\biggr)  + \frac{7}{18}\Lb
\biggr]\biggr\}, \\
& & 
\delta \gamma_1 =
\htt \biggl[ \muu
\biggl(-2 h_t^2(\Att+ |\hat\mu|^2)+
\fr43 g_S^2 +\fr34 g^2 \biggr) \Lb \nn
& & \hspace*{1cm}
+\fr14 m_{12}^2 B_6 (\hat A_t^*\hat\mu^*+\hat A_t\hat\mu)
(3 g^2-8 h_t^2 \muu) 
\biggr], \\
& & 
\delta \gamma_2 = 
\htt \biggl[
\biggl(
-2 h_t^2 (1-\Att)^2 -2 h_t^2 \Att\muu -\fr43 g_S^2 (1-\Att)\nn
& & \hspace*{0.5cm} 
- \fr34 g^2 (1-\Att) \biggr)\Lb 
+\fr14 m_{12}^2 B_6 (\hat A_t^*\hat\mu^*+\hat A_t\hat\mu)
(3 g^2 +8 h_t^2 (1-\Att))
\biggr], \hspace*{1cm} \\
& & 
\delta \gamma_{12} = 
\htt \biggl\{ \hat A_t^*\hat\mu^*
\biggl(
-2 h_t^2 (1-\Att-\muu) -\fr43 g_S^2 +
\fr34 g^2 \biggr) \Lb \nn
& & \hspace*{0.5cm}
+ \fr14 m_{12}^2 B_6
\Bigl[-3 g^2(1-\Att-\muu) +8 h_t^2 (\hat A_t^*\hat \mu^*)^2 
- 8 h_t^2 \muu (1-\Att)\Bigr]
\biggr\}, \\
& & 
\delta \lambda_1 =  
-\fr32 \htttt |\hat\mu|^4 \Lb
-\frac{g^4}{16\pi^2} \biggl[
\frac{9}{16} \biggl(\Lb - \fr23 \biggr)+ \fr14 \Lb
\biggr], \\
& & 
\delta \lambda_2 = 
-\fr32 \htttt (1-\Att)^2 \Lb
-\frac{g^4}{16\pi^2} \biggl[
\frac{9}{16} \biggl(\Lb - \fr23 \biggr)+ \fr14 \Lb
\biggr], \\
& & 
\delta \lambda_3 = 
 3 \htttt \muu (1-\Att) \Lb
-\frac{g^4}{16\pi^2} \biggl[
\frac{9}{8} \biggl(\Lb - \fr23 \biggr)+ \fr12 \Lb
\biggr], \\
& & 
\delta \lambda_4 = 
-3 \htttt \Att\muu \Lb
+\frac{g^4}{16\pi^2} \fr14 \Lb, \\
& & 
\delta \lambda_5 = 
-\frac{3}{16}\pf \biggl[
8 h_t^4 (\hat A_t^* \hat\mu^*)^2 \Lb + g^4 m_{12}^4 B_8 \biggr], \\
& & 
\delta \lambda_6 = 
\fr38 \pf \biggl[
8 h_t^4 \muu \hat A_t^*\hat\mu^* \Lb- g^4 m_{12}^2 B_6
\biggr], \\
& & 
\delta \lambda_7 = 
-\fr38 \pf \biggl[
8 h_t^4 (1-\Att) \hat A_t^*\hat\mu^* \Lb+ g^4 m_{12}^2 B_6
\biggr]. 
\ea

\subsection{Integrating out fermions}
\la{fermions}

The fermions to be integrated out, interacting with the 
Higgs and stop degrees of freedom, include 
the third generation quarks, the Higgsinos, 
and the SU(2) and SU(3) gauginos.
The mass parameters of the latter are denoted by $M_2,M_3$.

Concerning the integrals appearing, let us define
\ba
\Tint{p_f} \frac{1}{(p^2)^3} & = & \frac{7\zeta(3)}{128\pi^4T^2} \equiv 
\frac{F_6}{16\pi^2},\quad F_6 = \frac{7\zeta(3)}{8} \frac{1}{(\pi T)^2}, \\
\Tint{p_f} \frac{1}{(p^2)^4} & = & \frac{31\zeta(5)}{1024\pi^6T^4} \equiv 
\frac{F_8}{16\pi^2},\quad F_8 = \frac{31\zeta(5)}{64} \frac{1}{(\pi T)^4}, 
\ea
where $p_f$ denotes the fermionic Matsubara momenta.
Then the basic graphs give, in the 
high temperature expansion, contributions of the form

\vspace*{0.4cm}

\parbox[c]{2.5cm}{
\begin{picture}(70,40)(0,0)

\SetWidth{1.5}
\SetScale{0.9}
\DashLine(0,20)(20,20){5}
\DashLine(50,20)(70,20){5}
\CArc(35,20)(15,0,360)
\Text(2,28)[l]{$q$}

\end{picture}}
\parbox[c]{12.4cm}{
\ba
& & 
\Tint{p_f} \frac{\tr [i\pslash i(\,\,\,\,\slash \!\!\!\!\!\!\!\! p+q)]}
{[p^2+M_2^2][(p+q)^2+|\mu|^2]} \nn
& & =
\frac{T^2}{6}+ \pf(4M_2^2+4 |\mu|^2+2 {q}^2)\Lf,
\\
& & 
\Tint{p_f} \frac{1}{[p^2+M_2^2][(p+q)^2+|\mu|^2]} \nn
& & =\pf\biggl[
\Lf-\biggl(M_2^2+|\mu|^2+\fr13 {q}^2\biggr) F_6
\biggr], \hspace*{3.9cm}
\ea}



\parbox[c]{2.5cm}{
\begin{picture}(60,40)(0,0)

\SetWidth{1.5}
\SetScale{0.9}
\DashLine(0,0)(15,5){5}
\DashLine(0,40)(15,35){5}
\DashLine(45,5)(60,0){5}
\DashLine(45,35)(60,40){5}
\Line(15,5)(45,5)
\Line(45,5)(45,35)
\Line(15,5)(15,35)
\Line(15,35)(45,35)

\end{picture}}
\parbox[c]{12.4cm}{
\ba
& & 
\Tint{p_f} \frac{1}{[p^2+M_2^2][p^2+|\mu|^2]} 
\\
& & 
=\pf 
\biggl[\Lf - (M_2^2+|\mu|^2) F_6 + 
(M_2^4 + M_2^2|\mu|^2 + |\mu|^4) F_8
\biggr]. \nonumber
\ea}

\vspace*{0.4cm}

As the fermions appearing are Majorana particles, the
Lorentz-structure of the contractions is
\ba
& & 
\langle \widetilde{W}_\alpha(p) \overline{\widetilde{W}}_\beta(q) \rangle =
\delta(p-q) \frac{[-i\pslash+M]_{\alpha\beta}}{p^2+M^2}, \nn 
& & 
\langle \widetilde{W}_\alpha(p) \widetilde{W}_\beta(q) \rangle =
\delta(p+q) \frac{[(-i\pslash+M) C^{-1}]_{\alpha\beta}}{p^2+M^2}, \\
& & 
\langle \overline{\widetilde{W}}_\alpha(p) 
\overline{\widetilde{W}}_\beta(q) \rangle =
\delta(p+q) \frac{[C^{-1}(i\pslash+M)]_{\alpha\beta}}{p^2+M^2}, \nonumber
\ea
where the charge conjugation matrix $C$ satisfies
\be
C = -C^\dagger=-C^T=-C^{-1}, \quad
C^{-1} \gamma^\mu C = -(\gamma^\mu)^T, \quad
C^{-1} \gamma^5 C = (\gamma^5)^T,
\ee
and $\overline{\widetilde{W}}_\alpha(x) = 
\widetilde{W}_\beta(x) C_{\beta\alpha}$.

The final results for the couplings are (terms proportional to 
the gauge couplings come from loops involving gauginos, terms
proportional to the Yukawa coupling from loops involving
left-handed quarks):
\ba
& &
\delta Z_1 = \fr32 \gpf \Lf, \quad
\delta Z_2 = \biggl(\fr32 g^2+3 h_t^2
\biggr)\pf\Lf, \la{f1st} \\
& & 
\delta Z_{12} = -\gpf M_2 \mu^* F_6, \quad
\delta Z_U = \biggl(\fr83 g_S^2 + 2 h_t^2\biggr) \pf\Lf, \\
& & 
\delta m_1^2(T) = 
\fr34 g^2 \biggl(\frac{T^2}{6}+ 
4 (M_2^2 + |\mu|^2) \pf \Lf
\biggr), \la{fmm1} \\
& & 
\delta m_2^2(T) = 
\fr34 g^2 \biggl(\frac{T^2}{6}+ 
4(M_2^2 + |\mu|^2) \pf  \Lf
\biggr) + \fr14 h_t^2  T^2, \\
& &
\delta m_{12}^2(T) = 3 \gpf M_2 \mu^* \Lf, \\
& & 
\delta m_U^2(T) = \fr43 g_S^2 \biggl( \frac{T^2}{6} + 4 M_3^2 \pf\Lf \biggr)
+h_t^2 \biggl(\frac{T^2}{6} + 4 |\mu|^2 \pf\Lf  \biggr), 
\hspace*{1cm} \la{fmmU} \\
& & 
\delta \lambda_U = \pf\biggl(
2 h_t^4+ \frac{22}{9} g_S^4
\biggr)\Lf, \\
& & 
\delta \gamma_1 = -3 \gghtt |\mu|^2 F_6, \quad
\delta \gamma_2 =  \htt\biggl(
2 h_t^2 + \frac{16}{3} g_S^2 + 3 g^2 
\biggr) \Lf, \\
& & 
\delta \gamma_{12} = -3 \gghtt M_2 \mu^* F_6, \\ 
& & 
\delta \lambda_1 = \fr54 \gggg \Lf, \quad
\delta \lambda_2 = \pf\biggl( 
\fr54 g^4 + 3 h_t^4 \biggr)\Lf , \\
& & 
\delta \lambda_3 = \fr52 \gggg \Lf, \quad
\delta \lambda_4 = -2 \gggg \Lf, \\
& & 
\delta \lambda_5 = \fr32 \gggg M_2^2 (\mu^*)^2 F_8, \\
& & 
\delta \lambda_6 = -3 \gggg M_2 \mu^* F_6, \quad 
\delta \lambda_7 = -3 \gggg M_2 \mu^* F_6. \la{flast}
\ea

For completeness, let us also write down the
results for the CP-violating parameters in the 
limit that $M_2,|\mu|$ are large compared with $\pi T$
(this limit corresponds to the zero temperature
contributions considered in~\cite{mp}). This limit 
is of interest since it was seen in \eq\nr{regime}
that non-vanishing values of $M_2,|\mu|$ are preferred from
the point of view of spontaneous CP-violation, and it is 
thus important to know when the high-temperature expansion
used is reliable. We obtain
\ba
\delta Z_{12} & = &  
3 g^2 M_2 \mu^* \int\! dp\,
\frac{1}{(p^2+|\mu|^2)(p^2+M_2^2)^2}\biggl[
\fr43\frac{{\bf p}^2}{p^2+M_2^2}-1 \biggr]
\nn
& = & 
-\fr32 \frac{g^2}{16\pi^2} 
M_2 \mu^* \frac{|\mu|^4-M_2^4 + 4|\mu|^2 M_2^2 \ln\frac{M_2}{|\mu|}}
{(|\mu|^2-M_2^2)^3} \nn
&  \approx & -\fr12 \frac{g^2}{16\pi^2} \frac{\mu^*}{|\mu|}
\biggl[1+{\cal O}\Bigl(\frac{M_2^2-|\mu|^2}{M_2^2}\Bigr)^2\biggr],
\la{largeM2dZ12} \\
\delta m_{12}^2(T) & = & 
3 g^2 M_2 \mu^* \int\! dp\, \frac{1}{(p^2+M_2^2)(p^2+|\mu|^2)}
\nn
& = &
3 \gpf M_2 \mu^* \biggl[
\ln\frac{\bmu^2}{M_2 |\mu|}+1-\frac{M_2^2+|\mu|^2}{M_2^2-|\mu|^2}
\ln\frac{M_2}{|\mu|} \biggr] \nn
& \approx &
3 \gpf M_2 \mu^* \biggl[
\ln\frac{\bmu^2}{M_2^2} + 
{\cal O}\Bigl(\frac{M_2^2-|\mu|^2}{M_2^2}\Bigr)^2\biggr], 
\la{largeM2m12} \\
\delta \gamma_{12} & = &
-3g^2 h_t^2 M_2 \mu^* \int\! dp\, \frac{1}{(p^2+M_2^2)(p^2+|\mu|^2)^2}
\nn
& = &
-3 \gghtt \frac{M_2 \mu^*}{M_2^2-|\mu|^2}
\biggl[
2 \frac{M_2^2}{M_2^2-|\mu|^2}
\ln\frac{M_2}{|\mu|}-1
\biggr]  \nn
& \approx & 
-\fr32 \gghtt  \frac{\mu^*}{|\mu|}
\biggl[1+{\cal O}\Bigl(\frac{M_2^2-|\mu|^2}{M_2^2}\Bigr)^2\biggr],\\
\delta \lambda_5  & = & 
\fr32 g^4 M_2^2 (\mu^*)^2 
\int \! dp\, \frac{1}{(p^2+M_2^2)^2(p^2+|\mu|^2)^2}
\nn
& = &  3 \gggg \frac{M_2^2 (\mu^*)^2}{(M_2^2-|\mu|^2)^2}
\biggl[
\frac{M_2^2+|\mu|^2}{M_2^2-|\mu|^2}
\ln\frac{M_2}{|\mu|} -1
\biggr]  \nn
& \approx & 
\fr14\gggg \biggl(\frac{\mu^*}{|\mu|}\biggr)^2
\biggl[1+{\cal O}\Bigl(\frac{M_2^2-|\mu|^2}{M_2^2}\Bigr)^2\biggr],\la{l5t0} \\
\delta \lambda_6 & = & 
\delta \lambda_7 = 
-3 g^4 M_2 \mu^* \int\! dp\,\frac{p^2}{(p^2+M_2^2)^2(p^2+|\mu|^2)^2}
\nn
& = & 
-3 \gggg \frac{M_2 \mu^*}{(M_2^2-|\mu|^2)^3}
\biggl[ M_2^4-|\mu|^4 - 4 
M_2^2|\mu|^2
\ln\frac{M_2}{|\mu|}
\biggr]  \nn
& \approx & 
- \gggg \frac{\mu^*}{|\mu|}
\biggl[1+{\cal O}\Bigl(\frac{M_2^2-|\mu|^2}{M_2^2}\Bigr)^2\biggr],
\la{largeM2l6}
\ea 
where $dp=d^{4-2\epsilon}p/(2\pi)^{4-2\epsilon}$.
Comparing with the expressions in \eqs\nr{f1st}--\nr{flast}, 
we observe that the high-temperature expansion breaks down
when $\{M_2,\mu\}\gsim 2T$.
In the range $2T \lsim \{ M_2,|\mu| \} \lsim 4T$, 
a numerical evaluation of the integrals
appearing is needed, 
while at $\{ M_2,|\mu|\} \gsim 4T$, 
the values of the couplings saturate and
the zero-temperature 
results in \eqs\nr{largeM2dZ12}--\nr{largeM2l6} can be used. 
In \eqs\nr{largeM2dZ12}--\nr{largeM2l6} we have
shown also the direct integral representations which 
can easily be evaluated at any temperature
(see, e.g., \cite{fkot}). 

\subsection{Field redefinitions}
\la{redef}

In Secs.~\ref{mQ}--\ref{fermions}, the removal of various 
ultraviolet degrees of freedom from the theory resulted in 
contributions to the effective parameters 
characterizing the interactions of the infrared degrees 
of freedom. In addition, there were contributions to the
field normalizations,
$\delta Z_1,\delta Z_2,\delta Z_{12},\delta Z_U$. It is convenient
to make a redefinition of the fields so that  
these extra wave function terms disappear. This
results in new contributions to the effective parameters, 
rendering them gauge and scale independent at 1-loop level.

Summing the tree-level terms with those in \eq\nr{wavefunction}, 
the kinetic terms are of the form 
\ba
{\cal L}_\rmi{kin} & = &  
(D_i^w H_1)^\dagger (D_i^w H_1)(1+\delta Z_1) + 
(D_i^w H_2)^\dagger (D_i^w H_2)(1+\delta Z_2) \nn
& + & 
\Bigl[ (D_i^w H_1)^\dagger (D_i^w \tilde H_2) \delta Z_{12} + \Hc\Bigr] +
(D_i^s U)^\dagger (D_i^s U)(1+\delta Z_U).
\ea
One can now get rid of $\delta Z_1,...,\delta Z_U$ by writing 
\ba
H_1  & =  & \Bigl(1-\fr12 \delta Z_1\Bigr) H_1^\rmi{(new)} - 
\fr12 \delta Z_{12}\tilde H_2^\rmi{(new)}, \\
\tilde H_2 & = & 
- \fr12 \delta Z_{12}^* H_1^\rmi{(new)} +
\Bigl( 1- \fr12 \delta Z_2\Bigr) \tilde H_2^\rmi{(new)}, \\
U & = & \Bigl( 1- \fr12 \delta Z_U\Bigr) U^\rmi{(new)}.
\ea
When the field redefinitions combine with the tree-level couplings 
in \eq\nr{treelevel}, there are new contributions to the 
effective parameters as follows:
\ba
& &
\delta m_1^2(T) = -  m_1^2 \delta Z_1 -\fr12 m_{12}^2 (\delta Z_{12}+
\delta Z_{12}^*),  \\
& & 
\delta m_2^2(T) = -  m_2^2 \delta Z_2 -\fr12 m_{12}^2 
(\delta Z_{12}+\delta Z_{12}^*), \\
& & 
\delta m_{12}^2(T) = -\fr12  m_{12}^2 (\delta Z_1+\delta Z_2) 
-\fr12 (m_1^2+m_2^2) \delta Z_{12}, 
\\
& & 
\delta m_U^2(T) = - m_U^2 \delta Z_U, \quad
\delta \lambda_U = -\fr13 g_S^2 \delta Z_U, \\
& & 
\delta \gamma_1 = h_t^2 \muu (\delta Z_1+\delta Z_U)-
\fr12 h_t^2 (\delta Z_{12}^* 
\hat A_t^*\hat\mu^*+
\delta Z_{12} \hat A_t\hat\mu), \\
& & 
\delta \gamma_2 = -h_t^2 (1-\Att) (\delta Z_2+\delta Z_U) -
\fr12 h_t^2 (\delta Z_{12}^* \hat A_t^*\hat\mu^*+
\delta Z_{12} \hat A_t\hat\mu), \\
& & 
\delta \gamma_{12} = -\fr12  h_t^2(1-\Att-\muu) \delta Z_{12}-
\fr12 h_t^2 \hat A_t^*\hat\mu^*
(2 \delta Z_U + \delta Z_1+\delta Z_2) , \\
& & 
\delta \lambda_1 = - \fr14 g^2 \delta Z_1, \quad
\delta \lambda_2 = - \fr14 g^2 \delta Z_2, \\
& & 
\delta \lambda_3 = -\fr14 g^2 (\delta Z_1+\delta Z_2), \quad
\delta \lambda_4 = \fr12 g^2 (\delta Z_1+\delta Z_2).
\ea
There are no new contributions to $\lambda_5...\lambda_7$.

\subsection{Gauge couplings and $A_0,C_0$-integrations}
\la{gauge}

So far we have been considering couplings related to the 
Higgs fields and to the stop fields. Let us now review what
happens with the gauge fields. 

We start by inspecting the spatial sector,
where the plasma does not screen the gauge fields. 
What temperature does is that it changes the 
effective gauge coupling related to the static 
fields~\cite{generic,hl}:
\ba
g_T^2 & = & g^2(\bmu)\biggl\{
1+\frac{g^2}{48\pi^2}\biggl[
\beta_\rmi{gauge} \biggl(\ln\frac{\bmu}{\muT}+\frac{1}{22} \biggr)+
\beta_\rmi{scalar}\ln\frac{\bmu}{\muT} +
\beta_\rmi{fermion}\ln\frac{4\bmu}{\muT} \biggr]\biggr\}\nn
& = &
g_{T_0}^2 + \frac{g_{T_0}^4}{48\pi^2}(\beta_\rmi{gauge}+\beta_\rmi{scalar}+
\beta_\rmi{fermion})\ln\frac{T_0}{T},
\ea
where $T_0$ is any reference temperature, 
and the standard contributions to $\beta$ are,
for the weak gauge coupling $(N=2$), 
\ba
\beta_\rmi{gauge} & = &  22N, \\
\beta_\rmi{scalar} & = &  -N_H (\mbox{Higgses}) - 3 N_g (\mbox{squarks})
-N_g (\mbox{sleptons}), \\
\beta_\rmi{fermion} & = & -4N (\mbox{gauginos}) -2 N_H (\mbox{Higgsinos}) \nn
& & 
-6 N_g (\mbox{quarks}) -2 N_g  (\mbox{leptons}). 
\ea
Here $N_H=2$ is the number of Higgs doublets 
and $N_g=3$ the number of generations. 
For the superpartners, 
often only one generation (squarks) 
or none at all (sleptons) is taken to be 
effective. The expressions for the strong gauge coupling
are completely analogous (see also~\cite{bjls}).

Consider then the temporal components of the gauge fields. They do get 
screened. Neglecting higher order ${\cal O}(g^4)$-terms related
to $A_0,C_0$, the (tree-level) part of the Lagrangian involving
them is 
\ba
{\cal L}_\rmi{$A_0,C_0$} & = &
\fr12 (D_i^w A_0)^a(D_i^wA_0)^a + 
\fr12 (D_i^s C_0)^A(D_i^sC_0)^A + 
\fr12 m_{A_0}^2 A_0^aA_0^a + 
\fr12 m_{C_0}^2 C_0^AC_0^A \nn
& + &  
\fr14 g^2 A_0^aA_0^a (H_1^\dagger H_1+H_2^\dagger H_2)+
g_S^2 C_0^A C_0^B U_\alpha (T^A T^B)_{\alpha\beta} U_\beta^*.
\la{A0act}
\ea
Here the covariant derivatives are in the adjoint 
representation and the Debye masses are (see, e.g., \cite{coe})
\be
m_{A_0}^2 = \fr52 g^2 T^2, \quad
m_{C_0}^2 = \fr83 g_S^2 T^2, \la{debye}
\ee
where we took into account gauginos and Higgsinos 
($\delta m_{A_0}^2 = (1/2) g^2 T^2$), as well as
gluinos ($\delta m_{C_0}^2 = (1/2) g_S^2 T^2$), 
but only the 3rd generation $U$-squarks.

Now, usually the fields $A_0,C_0$ can be integrated out, 
since they have the mass scale $\sim gT$, while the phase
transition dynamics is related to the scale $\sim g^2T$. 
However, we have argued in \se\ref{sec:param}
that to study spontaneous CP-violation,
one has to keep also the Higgs degrees of freedom with masses
$\sim gT$ in the action. Thus it is, in terms of the original
4d power counting, not strictly speaking 
consistent to integrate out 
some degrees of freedom with masses $\sim gT$ and leave others
in the action. However, we can introduce 
a different power counting
within the 3d theory, in particular since numerically 
the Debye masses in \eq\nr{debye} are relatively large. 
Moreover, they do not affect the CP-violating couplings
at leading order. Thus it seems very reasonable to integrate
out also $A_0,C_0$. The 1-loop 
results of that integration are~\cite{bjls}:
\ba
& & 
g^{2{\rm (new)}} = 
g^2 \biggl(1-\frac{g^2 T}{24\pi m_{A_0}}
 \biggr), \quad
g_S^{2{\rm (new)}} = 
g_S^2 \biggl(1-\frac{g_S^2 T}{16\pi m_{C_0}} \biggr), \\
& & \delta m_1^2(T) = \delta m_2^2(T) = -\frac{3}{16\pi} g^2 T m_{A_0}, \quad
\delta m_U^2(T) = -\frac{1}{3\pi} g_S^2 T m_{C_0}, \\
& & 
\delta \lambda_U = -\frac{13}{36} \frac{g_S^4 T}{8\pi m_{C_0}}, \\
& & 
\delta \lambda_1 = 
\delta \lambda_2 = -\frac{3}{16} \frac{g^4 T}{8\pi m_{A_0}}, \quad 
\delta \lambda_3 = -\frac{3}{8} \frac{g^4 T}{8\pi m_{A_0}}.
\ea

\subsection{Vacuum renormalization}
\la{vacuum}

In the previous sections, we have derived an effective theory
in terms of renormalized parameters in the $\msbar$ scheme. In 
order to make connection to physics, the $\msbar$ scheme 
parameters should, in turn, be expressed in terms of 
measurable quantities. For completeness, we will discuss 
here some of the most important effects. A more detailed
discussion can be found, e.g., in~\cite{ml}--\cite{lo}.
In this section, we take, for simplicity, the couplings
$A_t,\mu$ to be real:
\be
A_t^* \to A_t, \quad
\mu^* \to \mu.
\ee
Of course, these quantities could still be negative. 
For the case of complex couplings see, e.g., \cite{ap}.

Let us note first that 
the problem with the renormalization of the gauge 
couplings $g_S^2, g^2$ and the top quark Yukawa coupling
$h_t$ is that the same parameters appear in many different
vertices due to supersymmetry, while in the softly broken
theory, this equivalence is lost beyond tree-level. Thus, 
it is not sufficient to compute three different physical
observables to fix these three parameters appearing in 
different places. However, as the renormalization 
effects related to these parameters are not too 
essential for the problem discussed in this paper, 
we shall not discuss the fixing of $g,h_t,g_S$ in great detail: 
it is sufficient to remark that the effective finite 
temperature gauge couplings in the gauge sector
are related to the zero temperature
$\msbar$ scheme couplings as discussed in \se\ref{gauge}, 
and the top quark Yukawa coupling can be fixed at 
tree-level, 
\be
h_t^2\ssb = \frac{g^2 m_t^2}{2 m_W^2},
\ee
where $\tan\beta=v_2/v_1$
is a parameter determined by the Higgs mass, see below.

Let us, instead, discuss the values of the 
mass parameters $m_1^2(\bmu), m_2^2(\bmu), m_{12}^2(\bmu)$, $m_U^2 (\bmu)$.
For them, there are relatively large renormalization effects proportional
to $h_t^2$, and we would like to account for
the dominant terms of such types, in particular those
proportional to $m_Q^2$.

The four running mass parameters can be fixed by computing
the 1-loop pole masses $m_Z^2,m_A^2,m_h^2, m_{\tilde t}^2$
of the Z boson, the CP-odd Higgs particle, the lightest CP-even 
Higgs particle, and the lightest stop, respectively. For the
dominant terms we are interested in, this computation can 
be simplified since the momentum dependence of the 2-point
functions can be neglected: indeed, the momentum dependence 
results essentially in multiplicative terms proportional to 
$p^2\sim m^2$, where $m^2$ is the 
pole mass in question. These terms are suppressed with respect 
to the dominant additive ones, $\sim m_Q^2$.
Since the momentum dependence can
be neglected, one can derive some of the 2-point 
functions from the effective potential.

There is one further simplification. It turns out that the only
large additive ($\sim m_Q^2$) contributions to the Z boson mass $m_Z^2$ 
come from the tadpole graphs which account for a shift in the 
location of the broken Higgs minimum $(v_1,v_2)$. In other
words, in the one-particle-irreducible graphs, 
the terms proportional to $m_Q^2$ cancel due to gauge invariance
(for explicit expressions
see, e.g., \eqs(A.23)--(A.25) in~\cite{ml}).
This means that if we denote by $(v_1,v_2)$ the location
of the radiatively corrected broken minimum, then we can write,
within the present approximation, 
$m_Z^2 = \tilde g^2(v_1^2+v_2^2)/4$, 
where $\tilde g^2 = g^2+g'^2$.
Then, all appearances of $v_1,v_2$
can be expressed in terms of $\tan\beta=v_2/v_1$ and $m_Z$,
as $\tilde g v_1= 2 m_Z \cos\!\beta$, 
$\tilde g v_2 = 2 m_Z \sin\!\beta$.

To solve for $m_1^2(\bmu),m_2^2(\bmu), m_{12}^2(\bmu)$, we now
proceed as follows. First, we solve from the conditions 
$\partial V(v_1,v_2)/\partial v_1=0, \partial V(v_1,v_2)/\partial v_2=0$
for $m_1^2(\bmu),m_2^2(\bmu)$ in terms of $\tan\beta, m_Z^2, 
m_{12}^2(\bmu)$. We feed these results into the mass matrix 
of the CP-odd Higgs (and of a Goldstone boson). Solving for
the physical eigenvalue $m_A^2$, we obtain $m_{12}^2(\bmu)$ as a function
of $\tan\beta, m_A^2, m_Z^2, \bmu$. With this expression, 
the results are known also for  $m_1^2(\bmu), m_2^2(\bmu)$. It remains
to eliminate $\tan\beta$: this is done by plugging the 
expressions found into the mass matrix of the CP-even Higgses, 
and thus finding the relation of $\tan\beta$ to the 
lightest CP-even Higgs mass $m_h$.

To proceed with this program, 
we write the Higgs doublets in component form as
\be
H_1=
\frac{1}{\sqrt{2}}
\left( \begin{array}{r} 
h_1^0 + i h_1^3 \\
-h_1^2 + i h_1^1
\end{array} \right), \quad
H_2=
\frac{1}{\sqrt{2}}
\left( \begin{array}{l} 
h_2^2 + i h_2^1  \\
h_2^0 - i h_2^3
\end{array} \right).
\la{h1h2}
\ee
The tree-level potential ($\< h_1^0\> =v_1, \< h_2^0 \> =v_2$) is then
\be
V^\rmi{tree}(v_1,v_2) = \fr12 m_1^2 v_1^2 + \fr12 m_2^2 v_2^2 + 
m_{12}^2 v_1v_2 + \frac{1}{32}\tilde g^2 (v_1^2-v_2^2)^2.
\ee
The $h_t^2$-contributions of 
the 1-loop potential to $\partial V/\partial v_i$
can be read, e.g., from \linebreak
\eqs(A.29), (A.30) in \cite{ml}:
\be
\frac{\partial V^\rmi{1-loop}(v_1,v_2)}{\partial v_1} = v_1 C_h(\bmu), \quad
\frac{\partial V^\rmi{1-loop}(v_1,v_2)}{\partial v_2} = v_2 S_h(\bmu),
\ee
where
\ba
& & C_h(\bmu) = 
3 \htt\biggl[
\mu (-\mu+A_t\tb)\biggl(
\ln\frac{\bmu^2}{m_{\tilde T}m_{\tilde t}} + 1 + 
\frac{m_{\tilde T}^2+m_{\tilde t}^2}{m_{\tilde T}^2-m_{\tilde t}^2}
\ln\frac{m_{\tilde t}}{m_{\tilde T}}
\biggr)
\biggr], \hspace*{1.5cm} \\
& & S_h(\bmu) = 
3 \htt\biggl[
2 m_\rmi{top}^2 \biggl(\ln\frac{\bmu^2}{m_\rmi{top}^2}+1\biggr) \nn
& & \hspace*{0.5cm} - 
(m_{\tilde t_L}^2+m_{\tilde t_R}^2)\biggl(
\ln\frac{\bmu^2}{m_{\tilde T}m_{\tilde t}} + 1\biggr)
-(m_{\tilde T}^2-m_{\tilde t}^2)
\ln\frac{m_{\tilde t}}{m_{\tilde T}} \nn
& & \hspace*{0.5cm} + A_t (-A_t+\mu\ctb)\biggl(
\ln\frac{\bmu^2}{m_{\tilde T}m_{\tilde t}} + 1 + 
\frac{m_{\tilde T}^2+m_{\tilde t}^2}{m_{\tilde T}^2-m_{\tilde t}^2}
\ln\frac{m_{\tilde t}}{m_{\tilde T}}
\biggr)
\biggr].
\ea
Here
\ba
& & m_{\tilde t_L}^2 =  m_Q^2 + m_\rmi{top}^2 + \fr12 m_Z^2 \cbb, \quad
m_{\tilde t_R}^2 = m_U^2 + m_\rmi{top}^2, \\
& & m_{\tilde t_{LR}}^2 = m_\rmi{top} (A_t - \mu\cot\beta), \\
& & m_{\tilde T}^2 = \fr12\Bigl[m_{\tilde t_L}^2 + m_{\tilde t_R}^2+
\sqrt{(m_{\tilde t_L}^2 - m_{\tilde t_R}^2)^2 + 4 m_{\tilde t_{LR}}^4} 
\Bigr] ,\\
& & m_{\tilde t}^2 = \fr12\Bigl[m_{\tilde t_L}^2 + m_{\tilde t_R}^2-
\sqrt{(m_{\tilde t_L}^2 - m_{\tilde t_R}^2)^2 + 4 m_{\tilde t_{LR}}^4} 
\Bigr].
\ea 
Note that in the notation of~\cite{ml}, 
$w_s \to -h_t\mu, u_s \to h_t A_t, e_s=d_s \to 0$, 
$m_{U1}^2\to m_{\tilde t_L}^2$, 
$m_{U2}^2\to m_{\tilde t_R}^2$,
$m_{U12}^2\to m_{\tilde t_{LR}}^2$,  
$m_{U+}^2 \to m_{\tilde T}^2$, 
$m_{U-}^2 \to m_{\tilde t}^2$. 
In addition, no Higgsino and gaugino contributions 
were considered in~\cite{ml}, but this does not affect 
the leading $h_t^2$ terms were are considering here
(see, however, the discussion below). 

It then follows from the conditions 
${\partial (V^\rmi{tree}+V^\rmi{1-loop})}/{\partial v_i} = 0$
that
\ba
m_1^2(\bmu) & = &  -m_{12}^2(\bmu)\tb-\fr12 m_Z^2 \cbb -C_h(\bmu), 
\la{m11m2} \\
m_2^2(\bmu) & = & -m_{12}^2(\bmu)\ctb + \fr12 m_Z^2\cbb -S_h(\bmu). \la{m1m2}
\ea

The mass matrix for the CP-odd Higgs mass $m_A$ and for 
a neutral Goldstone boson is given by the fields $h_1^3,h_2^3$, 
and is of the form 
\be
M_\rmi{CP-odd} = 
\left( \begin{array}{ll} 
m_1^2(\bmu) + \fr18 \tilde g^2 (v_1^2-v_2^2) + \Pi^A_{11}(\bmu) & 
m_{12}^2(\bmu) + \Pi^A_{12}(\bmu)  \\
m_{12}^2(\bmu) + \Pi^A_{12}(\bmu) &  
m_2^2(\bmu) + \fr18 \tilde g^2 (v_2^2-v_1^2) + \Pi^A_{22}(\bmu)
\end{array} \right) .
\ee
The expressions for $\Pi^A_{ij}(\bmu)$ can be obtained, e.g., 
from \eqs(A.19)--(A.22) in~\cite{ml} (the functions $F_H$ vanish
within the zero-momentum approximation considered, and the overall
sign should be reversed):
\ba
& & \Pi^A_{11}(\bmu) = C_h(\bmu) - \tb A_t \mu D(\bmu), \quad
\Pi^A_{12}(\bmu) = A_t \mu D(\bmu), \\
& & \Pi^A_{22}(\bmu) = S_h(\bmu) -\ctb A_t \mu D(\bmu), 
\ea
where
\be
D(\bmu) =  3\htt  \biggl(
\ln\frac{\bmu^2}{m_{\tilde T}m_{\tilde t}} + 1 + 
\frac{m_{\tilde T}^2+m_{\tilde t}^2}{m_{\tilde T}^2-m_{\tilde t}^2}
\ln\frac{m_{\tilde t}}{m_{\tilde T}}
\biggr).
\ee
Plugging in the values of $m_1^2(\bmu),m_2^2(\bmu)$ from 
\eqs\nr{m11m2}, \nr{m1m2}, the CP-odd mass matrix becomes 
\be
M_\rmi{CP-odd} = 
\left( \begin{array}{ll} 
-[m_{12}^2(\bmu)+ A_t \mu D(\bmu)]\tb & 
m_{12}^2(\bmu)+ A_t \mu D(\bmu) \\
m_{12}^2(\bmu)+ A_t \mu D(\bmu) & 
-[m_{12}^2(\bmu)+ A_t \mu D(\bmu)]\ctb
\end{array} \right). 
\ee
The (scale-independent) eigenvalues of $M_\rmi{CP-odd}$ are
zero and $m_A^2$, which finally gives the expression for 
$m_{12}^2(\bmu)$, and after insertion into \eqs\nr{m11m2}, \nr{m1m2}, 
for $m_1^2(\bmu), m_2^2(\bmu)$:
\ba
& & m_{12}^2(\bmu) = -\fr12 m_A^2\sbb - A_t \mu D(\bmu), \la{T0m12} \\
& & m_1^2(\bmu) = \fr12 m_A^2 -\fr12 (m_A^2+m_Z^2) \cbb + 
|\mu|^2 D(\bmu) , \la{mm1} \\
& & m_2^2(\bmu) = \fr12 m_A^2 + \fr12 (m_A^2+m_Z^2) \cbb + 
|A_t|^2 D(\bmu) \la{finalmasses} \\
& & \hspace*{0.5cm} + 3 \htt \biggl[
(m_Q^2+m_U^2)  
\biggl(
\ln\frac{\bmu^2}{m_{\tilde T}m_{\tilde t}} + 1 \biggr) + 
2 m_\rmi{top}^2 \ln\frac{m_\rmi{top}^2}{m_{\tilde t} m_{\tilde T}}
+ (m_{\tilde T}^2-m_{\tilde t}^2)
\ln\frac{m_{\tilde t}}{m_{\tilde T}}
\biggr].  \hspace*{0.4cm}\nonumber
\ea

There is the following useful observation to be made. If one 
expands \eqs\nr{T0m12}--\nr{finalmasses} in terms of 
$m_{\tilde t}^2/m_Q^2, m_\rmi{top}^2/m_Q^2$
in the limit of large $m_Q$, then it turns out that the 
dominant terms containing $m_Q$ cancel against the terms in 
\eqs\nr{mQmm1}--\nr{mQm12}. For example, summing the
dominant 1-loop contributions to $m_{12}^2(T)$
from \eqs\nr{mQm12}, \nr{bosonm12} together with \eq\nr{T0m12}, 
the scale dependence cancels and we obtain, 
up to terms of relative order $m_{\tilde t}^4/m_Q^4, m_\rmi{top}^4/m_Q^4$,
\be
\delta m_{12}^2(T) = 
3 \htt \hat A_t^*\hat \mu^*\biggl\{
m_U^2 \biggl(\ln\frac{\muT^2}{m_{\tilde t}^2}+1 \biggr) 
+ m_\rmi{top}^2 \biggl[ 
1+\frac{\tilde A_t^2}{m_Q^2}+
2 \biggl(1-\frac{\tilde A_t^2}{m_Q^2}\biggr) \ln\frac{m_Q}{m_{\tilde t}}
\biggr]\biggr\}, \la{remnant} 
\ee
where we have denoted 
$m_{\tilde t_{LR}}^2 = m_\rmi{top} \tilde A_t$. 
The largest term here is that induced by the symmetry
breaking, $\sim A_t^*\mu^* (m_\rmi{top}/m_Q)^2$. For 
$m_1^2(T),m_2^2(T)$, in particular (where the result is similar), 
such terms are typically much smaller than the tree-level mass terms. 
Thus, after vacuum renormalization is taken into account, the terms 
proportional to $h_t^2/(16\pi^2)$ in $m_1^2(T),m_2^2(T)$ can
be neglected as a first approximation, and it is sufficient
to consider the tree-level mass terms and the thermal $T^2$-terms. 

By similar arguments, we can also easily derive
the dominant contributions of the parameters $M_2,|\mu|$ in the 
limit that $M_2,|\mu|\gg m_W$.
Indeed, in this limit the expressions for vacuum renormalization
must cancel the large direct contributions in \eq\nr{largeM2m12},
just as happened above for $m_Q$
in the limit $m_Q\gg m_\rmi{top}$. Thus, e.g., 
\be
m_{12}^2(\bmu) \approx 
\left. m_{12}^2(\bmu) \right|_{\mbox{\nr{T0m12}}}
-\left.
\delta m_{12}^2 (T) \right|_{\mbox{\nr{largeM2m12}}}. \la{m12ren}
\ee

Finally, let us connect $\tb$ to the Higgs mass.
The mass matrix for the CP-even Higgses $m_h,m_H$ 
is given by the fields $h_1^0,h_2^0$: 
\be
M_\rmi{CP-even} = 
\left( \begin{array}{ll} 
m_1^2(\bmu) + \fr18 \tilde g^2 (3 v_1^2-v_2^2) + \Pi^H_{11}(\bmu) & 
m_{12}^2(\bmu)-\fr14\tilde g^2v_1v_2 + \Pi^H_{12}(\bmu) \\
m_{12}^2(\bmu)-\fr14\tilde g^2v_1v_2 + \Pi^H_{12}(\bmu) & 
m_2^2(\bmu) + \fr18 \tilde g^2 (3 v_2^2-v_1^2) + \Pi^H_{22}(\bmu)
\end{array} \right), \la{evenM}
\ee
where the $\Pi^H_{ij}(\bmu)$'s can be obtained from the second 
derivatives of the effective potential $V(v_1,v_2)$, or 
directly from a diagrammatic computation in, e.g., 
\eqs(A.12)--(A.18) in~\cite{ml}. Denoting
\ba
\delta_1 & = & 12 \htt \biggl(
1+\frac{m_{\tilde T}^2+m_{\tilde t}^2}{m_{\tilde T}^2-m_{\tilde t}^2}
\ln\frac{m_{\tilde t}}{m_{\tilde T}} \biggr)
\frac{|\tilde A_t|^2 m_\rmi{top}^2}{(m_{\tilde T}^2-m_{\tilde t}^2)^2}, \\
\delta_2 & = & 12 \htt \frac{m_\rmi{top}^2}{m_{\tilde T}^2-m_{\tilde t}}
\ln \frac{m_{\tilde T}}{m_{\tilde t}}, \\
\delta_3 & = & 12 \htt \ln \frac{m_{\tilde t} m_{\tilde T}}{m_\rmi{top}^2},
\ea
where again $\tilde A_t=A_t- \mu\cot\!\beta$, 
we obtain
\ba
& & \Pi^H_{11}(\bmu) = -|\mu|^2 \Bigl[D(\bmu)-\delta_1\Bigr], \quad
\Pi^H_{12}(\bmu) = A_t \mu \Bigl[D(\bmu)-\delta_1\Bigr]
-\tilde A_t \mu \delta_2, \\
& & \Pi^H_{22}(\bmu) = -|A_t|^2 \Bigl[D(\bmu)-\delta_1\Bigr]+
2\tilde A_t A_t \delta_2 + \fr32  m_\rmi{top}^2 \delta_3 \nn
& & \hspace*{1cm} - 3 \htt \biggl[
(m_Q^2+m_U^2)\biggl(
\ln\frac{\bmu^2}{m_{\tilde T}m_{\tilde t}} + 1\biggr)   
+(m_{\tilde T}^2-m_{\tilde t}^2)
\ln\frac{m_{\tilde t}}{m_{\tilde T}}\biggr].
\ea

Plugging these expressions together with those
in \eqs\nr{T0m12}--\nr{finalmasses} into \linebreak \eq\nr{evenM}, 
the explicit scale dependence cancels and the 
physical mass matrix remaining is (see also~\cite{erz})
\ba
M_\rmi{CP-even} & = &  
\left( \begin{array}{ll} 
M_{11} & M_{12} \\
M_{12}^* & M_{22} 
\end{array} \right), \\
M_{11} & = & m_Z^2\ccb+m_A^2\ssb+|\mu|^2\delta_1, \\
M_{12} & = & -(m_Z^2+m_A^2) \sb\cb-A_t\mu \delta_1
-\tilde A_t\mu \delta_2, \\
M_{22} & = & m_Z^2 \ssb+m_A^2 \ccb + 
|A_t|^2 \delta_1 + 2 \tilde A_t A_t \delta_2+ m_\rmi{top}^2 \delta_3.
\ea
The lightest CP-even Higgs mass is then
\ba
m_h^2 & = &  \fr12 \biggl[M_{11}+M_{22} -  
\sqrt{(M_{11}-M_{22})^2 + 4 |M_{12}|^2}
\biggr].
\ea

The renormalization of $m_U^2(\bmu)$ and its relation to the 
lightest stop mass $m_{\tilde t}$ (which, in the absence of mixing,
is the right-handed stop mass $m_{\tilde t_R}$) can be handled 
with similar methods as the Higgs sector. However, we have already 
argued what the result will approximately be:  there are large
renormalization effects related to the integration out 
of $Q$, see \eq\nr{mQmU}, but when vacuum renormalization 
is taken into account, then these effectively cancel.
Thus, it is enough for our purposes to ignore
the effects in \eq\nr{mQmU}
and replace $m_U^2$ with the tree-level value, which is 
obtained from the expression for the lightest stop mass: 
\be
m_{\tilde t}^2 \approx m_U^2 + 
m_\rmi{top}^2 \Bigl[
1 - (\hat A_t-\hat\mu\cot\beta)^2
\Bigr].
\ee

\subsection{Summary: the effective theory}
\la{summary}

In this section we collect together the results of the 
previous sections. For the quartic couplings, 
we will work at the accuracy of the vacuum 
renormalization discussed in \se\ref{vacuum}. Thus, we ignore 
all the contributions of order ${\cal O}(g^4)$ to the quartic 
Higgs couplings, contributions of order ${\cal O}(g_S^4)$
to the quartic stop coupling, and contributions of order
${\cal O}(h_t^4)$ to the top Yukawa coupling squared appearing
in the quartic scalar interactions between stops and Higgses.
In other words, we keep only contributions for which the scale 
dependence is cancelled within the accuracy of vacuum 
renormalization discussed in \se\ref{vacuum}.
It should be noted, though, that for the quartic stop coupling,
neglecting effects of order ${\cal O}(g_S^4)$ and keeping
effects of order ${\cal O}(h_t^4)$ is  
numerically not a very useful approximation scheme. 

We also make an expansion in $m_i^2/m_Q^2, m_i^2/(\pi T)^2$,
where $m_i^2$ denotes the mass parameters other than $m_Q$.  
In general, we keep the dominant effects of the different
qualitative types appearing. Note that $|A_t|$ can, in principle,
be larger than $|\mu|$, since it is not subject to constraints
from the high-temperature expansion. The Debye masses 
appearing are given in \eq\nr{debye}. 

For the mass parameters $m_1^2(T),m_2^2(T), m_U^2(T)$, 
we only keep the leading terms \linebreak below, which should already 
give a reasonable approximation as explained after \linebreak 
\eq\nr{remnant}. More precise results can be obtained by using 
the full expressions in \eqs\nr{mm1},\nr{finalmasses} 
(see also~\cite{ml}--\cite{lo}).

For $m_{12}^2(T)$, which affects directly spontaneous CP-violation, 
we display a more precise expression, with the dominant 
additive corrections of various qualitative types included. 
Multiplicative corrections to $m_A^2\sbb$
have been neglected. It should also be noted that  
the high temperature expansion  for the gaugino-Higgsino
contribution is reliable
only when the expression in the square brackets on the 
first row in \eq\nr{seq3rd} is negative: 
without a high temperature expansion, 
this term is
\be
\delta m_{12}^2(T) = 
3 g^2 \frac{M_2\mu^*}{M_2^2-|\mu|^2}
\biggl[
I_f\Bigl(\frac{|\mu|}{T}\Bigr) - 
I_f\Bigl(\frac{M_2}{T}\Bigr) \biggr], \la{gHm12}
\ee 
where
\be
I_f(y) = -\frac{T^2}{2\pi^2} \int_0^{\infty} \! dx\, x^2 
\frac{1}{\sqrt{x^2+y^2}}
\frac{1}{e^{\sqrt{x^2+y^2}}+1}.
\ee
If $M_2, |\mu|$ are large ($\gsim \pi T$), then the 
term in \eq\nr{gHm12} vanishes, but
also the contributions of these degrees of freedom to the 
$T^2$-terms in the mass parameters, \eqs\nr{fmm1}--\nr{fmmU},
and to the Debye masses, \eq\nr{debye}, should be left out. 

Note that the square bracket on the second row
in \eq\nr{seq3rd} is also negative, so that the whole term 
is positive for $\mathop{\rm sign}(\hat A_t^*\hat\mu^*) > 0$; 
see \eq\nr{remnant} for an approximation. 

Finally, let us point out that we write here the couplings
in 4d units, so that there is an overall factor $\int_0^\beta d\tau=1/T$ 
in the action. This factor can, as usual, be defined away by 
a rescaling of the fields, and the result is that the quartic
couplings of the 3d action get simply multiplied by $T$.

With these conventions and
approximations, we obtain that\footnote{Note that 
as can be seen from the $h_t^4$-term in \eq\nr{finl2}, 
there is an error in \eq(3.7) of the latter of~\cite{mssmsim}:
there should be no extra factor of $e^{3/4}$ inside the logarithm.
(This does not affect any of the conclusions in~\cite{mssmsim}
concerning the non-perturbative effects.)}
\ba
m_1^2(T) & = &  \fr12 m_A^2 - \fr12 (m_A^2+m_Z^2)\cbb \nn
& & + 
\biggl(\fr38 g^2 -\fr14 h_t^2 \muu  \biggr) T^2
-\frac{3}{16\pi} g^2 T m_{A_0}, \la{seq1st} \\
m_2^2(T) & = &  \fr12 m_A^2 + \fr12 (m_A^2+m_Z^2)\cbb \nn
& &  + 
\biggl(\fr38 g^2+\fr12 h_t^2 -\fr14 h_t^2 \Att  \biggr) T^2
-\frac{3}{16\pi} g^2 T m_{A_0}, \la{seq2nd} \\
m_{12}^2(T) & = &  -\fr12 m_A^2\sbb + 
3 \gpf M_2\mu^* \biggl[\ln\frac{16 M_2|\mu|}{\muT^2}-1+
\frac{M_2^2+|\mu|^2}{M_2^2-|\mu|^2}\ln\frac{M_2}{|\mu|} \biggr] \nn
& & 
- 3\htt A_t^* \mu^* 
\biggl[
\ln\frac{m_Q^2}{m_{\tilde T}^2} - 
2 \frac{m_{\tilde t}^2}{m_{\tilde T}^2-m_{\tilde t}^2}
\ln\frac{m_{\tilde T}}{m_{\tilde t}} + 
2 \frac{m_U^2}{m_Q^2} \ln\frac{m_Q}{e^{1/2} \muT}
\biggr] \nn
& & 
+ m_A^2\pf\biggl(
\fr34 h_t^2 \hat A_t^*\hat\mu^* + \frac{7\zeta(3)}{16} g^2 
\frac{M_2\mu^*}{(\pi T)^2} \biggr)
+ 
\fr14 h_t^2 \hat A_t^*\hat\mu^* T^2, \la{seq3rd} \\
m_U^2(T) & = &  
m_{\tilde t}^2 - m_\rmi{top}^2 \Bigl[
1-|\hat A_t - \hat\mu^* \cot\beta|^2 \Bigr] \nn
& & + 
\biggl( \fr23 g_S^2 + \fr13 h_t^2 -\fr16 h_t^2 (\Att+\muu)
\biggr) T^2
-\frac{1}{3\pi} g_S^2 T m_{C_0}, \\
\lambda_U & = & \fr16 g_S^2 -\frac{13}{36} \frac{g_S^4 T}{8\pi m_{C_0}}
+ 
\fr13 \htt \biggl[ 
-g_S^2 \biggl(4\lnfQT+3\Att+3\muu \biggr) \nn
& & 
+ 6 h_t^2 \biggl(
\lnstQT+2 \Att\lnQT+\muu-(\Att+\muu)^2\lnQeT 
\biggr) \biggr], \\
\gamma_1 & = &   -h_t^2 \muu 
-\frac{\zeta(3)}{4}
\gghtt
\biggl[-\fr12 
\frac{m_A^2\sbb}{(2\pi T)^2} (\hat A_t^*\hat\mu^* + \hat A_t\hat \mu)  \nn
& &  + 
7 \frac{|\mu|^2}{(2\pi T)^2}\biggl(
6-\frac{M_2}{m_Q} (\hat A_t^*+\hat A_t) 
\biggr)
\biggr], \\
\gamma_2 & = &  
h_t^2 (1-\Att) , \\ 
\gamma_{12} &  =  & 
h_t^2 \hat A_t^* \hat \mu^*  
-\frac{\zeta(3)}{4} \htt
\biggl[
-2 \frac{m_A^2\sbb}{(2\pi T)^2} \Bigl(
g^2 (1-\Att-\muu) \nn
& & + h_t^2 \muu (1-\Att) \Bigr) 
+ 7 g^2 \frac{M_2 \mu^*}{(2\pi T)^2} (5+\Att+\muu) \biggr], \\
\lambda_1 & = &  
\fr18 (g^2+g'^2) -\frac{3}{16} \frac{g^4 T}{8\pi m_{A_0}} 
+ \fr38 \htt \muu \biggl( 
g^2 - 8 h_t^2 \muu \lnQeT \biggr), \\
\lambda_2 & = &  
\fr18 (g^2+g'^2)-\frac{3}{16} \frac{g^4 T}{8\pi m_{A_0}} 
+ 
\fr38 \htt \biggl[
-g^2 \biggl(4\lnfQT + 3 \Att \biggr) 
\hspace*{2.0cm} \nn 
& &
+ 8 h_t^2 \biggl(
\lnstQT+ 2\Att \lnQT- |\hat A_t|^4 \lnQeT \biggr) \biggr], \la{finl2} \\
\lambda_3 & = &
\fr14 (g^2-g'^2) -\frac{3}{8} \frac{g^4 T}{8\pi m_{A_0}}
+\fr38 \htt \biggl[
-g^2  \biggl(4\lnfQT + 3 \Att -\muu \biggr)
\nn 
& & 
+ 16 h_t^2 \muu \biggl(
\lnQheT-\Att\lnQeT \biggr)
\biggr], \\
\lambda_4 & = & 
-\fr12 g^2 
+\fr34 \htt \biggl[
g^2  \biggl(4\lnfQT + 3 \Att -\muu \biggr)
\nn
& & 
+ 4 h_t^2 \muu \biggl(
1-2 \Att\lnQeT \biggr)
\biggr], \\
\lambda_5 & = & 
\pf \biggl[
-3 h_t^4 (\hat A_t^*\hat \mu^*)^2 \lnQeT \nn
& & - 
\frac{3\zeta(5)}{64} g^4 \biggl( \fr14 \frac{m_A^4\sin^2 2\beta}{(2\pi T)^4} - 
\frac{31}{2} \frac{(M_2\mu^*)^2}{(\pi T)^4} \biggr)
\biggr], \la{seql5} \\
\lambda_6 & = & 
\pf \biggl[
-\fr34 h_t^2 \hat A_t^* \hat\mu^* 
\biggl(g^2 - 8 h_t^2 \muu \lnQeT \biggr) \nn
& & -
\frac{3\zeta(3)}{16} g^4 \biggl( -\fr12 \frac{m_A^2\sbb}{(2\pi T)^2} +
14 \frac{M_2 \mu^*}{(\pi T)^2}\biggr)
\biggr], \\
\lambda_7 & = & 
\pf \biggl[
\fr34 h_t^2 \hat A_t^* \hat\mu^* 
\biggl(g^2 - 8 h_t^2 \lnQT + 8 h_t^2\Att \lnQeT \biggr) \nn
& & 
- \frac{3\zeta(3)}{16} g^4 \biggl( -\fr12 \frac{m_A^2\sbb}{(2\pi T)^2} +
14 \frac{M_2 \mu^*}{(\pi T)^2} \biggr)
\biggr]. \la{seqlast}
\ea
In these formulas, all the scale dependence has cancelled.



\begin{thebibliography}{99}

\bibitem{krs}
V.A. Kuzmin, V.A. Rubakov and M.E. Shaposhnikov,
Phys.\ Lett.\ B 155 (1985) 36;
M.E. Shaposhnikov, Nucl.\ Phys.\ B 287 (1987) 757.
 
\bibitem{rs}
V.A. Rubakov and M.E. Shaposhnikov,
Usp.\ Fiz.\ Nauk 166 (1996) 493 [hep-ph/9603208]. 


\bibitem{own}
M. Laine and K. Rummukainen, review presented at 
{\em Lattice '98} [hep-lat/9809045].


\bibitem{e}
J.R. Espinosa, Nucl.\ Phys.\ B 475 (1996) 273 [hep-ph/9604320].

\bibitem{ce}
B. de Carlos and J.R. Espinosa,
Nucl.\ Phys.\ B 503 (1997) 24 [hep-ph/9703212].

\bibitem{bjls}
D. B\"odeker, P. John, M. Laine and M.G. Schmidt, 
Nucl.\ Phys.\ B 497 (1997) 387 [hep-ph/9612364].

\bibitem{cqw2}
M. Carena, M. Quir\'os and C.E.M. Wagner,
Nucl.\ Phys.\ B 524 (1998) 3 [hep-ph/9710401].

\bibitem{cm}
J.M. Cline and G.D. Moore, 
Phys.\ Rev.\ Lett.\ 81 (1998) 3315 [hep-ph/9806354];
J.M. Cline, 
McGill-98/27 [hep-ph/9810267]. 


\bibitem{ml}
M. Laine, Nucl.\ Phys.\ B 481 (1996) 43 [hep-ph/9605283];
Nucl.\ Phys.\ B, in press (E).

\bibitem{ck}
J.M. Cline and K. Kainulainen, 
Nucl.\ Phys.\ B 482 (1996) 73 [hep-ph/9605235];
Nucl.\ Phys.\ B 510 (1998) 88 [hep-ph/9705201].

\bibitem{lo}
M. Losada, 
Phys.\ Rev.\ D 56 (1997) 2893 [hep-ph/9605266];
G.R. Farrar and M. Losada, 
Phys.\ Lett.\ B 406 (1997) 60 [hep-ph/9612346];
M. Losada, 
Nucl.\ Phys.\ B 537 (1999) 3 
[hep-ph/9806519].

\bibitem{mssmsim}
M. Laine and K. Rummukainen, 
Phys.\ Rev.\ Lett.\ 80 (1998) 5259 [hep-ph/9804255];
Nucl.\ Phys.\ B 535 (1998) 423 [hep-lat/9804019].

\bibitem{asy}
P. Arnold, D.T. Son and L.G. Yaffe, 
Phys.\ Rev.\ D 55 (1997) 6264 [hep-ph/9609481];
UW-PT-98-10 [hep-ph/9810216].

\bibitem{db}
D. B\"odeker, 
Phys.\ Lett.\ B 426 (1998) 351 [hep-ph/9801430];
HD-THEP-98-35 [hep-ph/9810265].

\bibitem{moore_sy}
G.D. Moore, 
McGill-98-28 [hep-ph/9810313],
and references therein. 

\bibitem{moore_br}
G.D. Moore, 
Phys.\ Lett.\ B 439 (1998) 357
[hep-ph/9801204]; 
Phys.\ Rev.\ D 59 (1999) 014503
[hep-ph/9805264].


\bibitem{wb}
W. Bernreuther, 
lectures at the 37th International School of Particle Physics, 
Schladming, Austria, 1998 [hep-ph/9808453]. 

\bibitem{bgk}
T. Falk and K.A. Olive, 
Phys.\ Lett.\ B 439 (1998) 71 [hep-ph/9806236];
T. Ibrahim and P. Nath, 
Phys.\ Rev.\ D 58 (1998) 111301 [hep-ph/9807501];
M. Brhlik, G.J. Good and G.L. Kane, 
UM-TH-98-16 [hep-ph/9810457];
D. Chang, W.-Y. Keung and A. Pilaftsis, 
CERN-TH/98-343 [hep-ph/9811202];
and references therein.


\bibitem{lee}
T.D. Lee, 
Phys.\ Rev.\ D 8 (1973) 1226.

\bibitem{mp}
N. Maekawa, 
Phys.\ Lett.\ B 282 (1992) 387;
A. Pomarol, 
Phys.\ Lett.\ B 287 (1992) 331.


\bibitem{emq}
D. Comelli and M. Pietroni,
Phys.\ Lett.\ B 306 (1993) 67 [hep-ph/9302207];
J.R. Espinosa, J.M. Moreno and M. Quir\'os,
Phys.\ Lett.\ B 319 (1993) 505 [hep-ph/9308315].

\bibitem{cpr}
D. Comelli, M. Pietroni and A. Riotto,
Nucl.\ Phys.\ B 412 (1994) 441 [hep-ph/9304267]. 

\bibitem{fkot}
K. Funakubo, A. Kakuto, S. Otsuki and F. Toyoda,
Prog.\ Theor.\ Phys.\ 98 (1997) 427 [hep-ph/9704359]; 
Prog.\ Theor.\ Phys.\ 99 (1998) 1045 [hep-ph/9802276]. 


\bibitem{mstv}
L. McLerran, M. Shaposhnikov, N. Turok and M. Voloshin, 
Phys.\ Lett.\ B 256 (1991) 451.

\bibitem{nck}
A.E. Nelson, D.B. Kaplan and A.G. Cohen, 
Nucl.\ Phys.\ B 373 (1992) 453.

\bibitem{jpt}
M. Joyce, T. Prokopec and N. Turok, 
Phys.\ Rev.\ D 53 (1996) 2930 [hep-ph/9410281];
Phys.\ Rev.\ D 53 (1996) 2958 [hep-ph/9410282].

\bibitem{hn}
P. Huet and A.E. Nelson, 
Phys.\ Rev.\ D 53 (1996) 4578 [hep-ph/9506477].

\bibitem{ckv}
J.M. Cline, K. Kainulainen and A.P. Vischer, 
Phys.\ Rev.\ D 54 (1996) 2451 [hep-ph/9506284].

\bibitem{non-eq}
M. Carena, M. Quir\'os, A. Riotto, I. Vilja and C.E.M. Wagner,
Nucl.\ Phys.\ B 503 (1997) 387 [hep-ph/9702409].

\bibitem{cjk}
J.M. Cline, M. Joyce and K. Kainulainen,
Phys.\ Lett.\ B 417 (1998) 79 [hep-ph/9708393].

\bibitem{rio}
A. Riotto, 
Phys.\ Rev.\ D 58 (1998) 095009 [hep-ph/9803357], 
and references therein.

\bibitem{kl}
H. Kurki-Suonio and M. Laine,
Phys.\ Rev.\ Lett.\ 77 (1996) 3951 [hep-ph/9607382].

\bibitem{generic} 
K. Kajantie, M. Laine, K. Rummukainen and M. Shaposhnikov,
Nucl.\ Phys.\ B 458 (1996) 90 [hep-ph/9508379].

\bibitem{cfh}
F. Csikor, Z. Fodor and J. Heitger, 
Phys.\ Rev.\ Lett.\ 82 (1999) 21
[hep-ph/9809291].

\bibitem{su2u1} 
K. Kajantie, M. Laine, K. Rummukainen and M. Shaposhnikov,
Nucl. Phys. B 493 (1997) 413 [hep-lat/9612006].

\bibitem{pviol}
K. Kajantie, M. Laine, K. Rummukainen and M. Shaposhnikov,
Phys.\ Lett.\ B 423 (1998) 137 [hep-ph/9710538].

\bibitem{clm}
J.A. Casas, A. Lleyda and C. Mu\~noz, 
Nucl.\ Phys.\ B 471 (1996) 3 [hep-ph/9507294]. 

\bibitem{bgg}
G.C. Branco, J.-M. Gerard and W. Grimus, 
Phys.\ Lett.\ B 136 (1984) 383.

\bibitem{mqs}
J.M. Moreno, M. Quir\'os and M. Seco,
Nucl.\ Phys.\ B 526 (1998) 489 [hep-ph/9801272].

\bibitem{pj}
P. John, 
HD-THEP-98-54 [hep-ph/9810499].


\bibitem{nmssm}
D. Comelli, M. Pietroni and A. Riotto, 
Phys.\ Lett.\ B 343 (1995) 207 [hep-ph/9410225].


\bibitem{rt}
A. Riotto and O. T\"ornkvist, 
Phys.\ Rev.\ D 56 (1997) 3917 [hep-ph/9704371].


\bibitem{adjoint}
K. Kajantie, M. Laine, A. Rajantie, K. Rummukainen and M. Tsypin,
J.\ High Energy Phys.\ 11 (1998) 011 [hep-lat/9811004].


\bibitem{rosiek}
J. Rosiek,
Phys.\ Rev.\ D 41 (1990) 3464; hep-ph/9511250 (E).

\bibitem{joa}
J.O. Andersen, OHSTPY-HEP-T-98-007 [hep-ph/9804280].

\bibitem{ceqw}
M. Carena, J.R. Espinosa, M. Quir\'os and C.E.M. Wagner, 
Phys.\ Lett.\ B 355 (1995) 209 [hep-ph/9504316].

\bibitem{hl}
S. Huang and M. Lissia, 
Nucl.\ Phys.\ B 438 (1995) 54 [hep-ph/9411293].


\bibitem{coe}
D. Comelli and J.R. Espinosa, 
Phys.\ Rev.\ D 55 (1997) 6253 [hep-ph/9606438].





\bibitem{ap}
A. Pilaftsis, 
Phys.\ Rev.\ D 58 (1998) 096010 [hep-ph/9803297];
Phys.\ Lett.\ B 435 (1998) 88 [hep-ph/9805373].

\bibitem{erz}
J. Ellis, G. Ridolfi and F. Zwirner, 
Phys.\ Lett.\ B 262 (1991) 477.

\end{thebibliography}
\end{document}